\documentclass[twoside]{IEEEtran}

\usepackage{algorithm}
\usepackage{algpseudocode}
\usepackage{amsmath}
\usepackage{amssymb}
\usepackage{amsthm}
\usepackage{array}
\usepackage{balance}
\usepackage{blindtext}
\usepackage{booktabs}
\usepackage{caption}
\usepackage{color}
\usepackage{colortbl}
\usepackage{graphicx}
\usepackage{hhline}
\usepackage{lipsum}
\usepackage{mathtools}
\usepackage{multirow}
\usepackage{relsize}
\usepackage{soul}
\usepackage{stackengine}
\usepackage{subfiles}
\usepackage{textcase}
\usepackage{textcomp}
\usepackage{tikz}
\usepackage{xcolor}

\usepackage{hyperref}
\hypersetup{
colorlinks=true,
linkcolor=blue,
filecolor=magenta,      
urlcolor=blue,
}

\definecolor{white} {rgb}{1.00,1.00,1.00}
\definecolor{lgray} {rgb}{0.95,0.95,0.95}
\definecolor{gray}  {rgb}{0.90,0.90,0.90}
\definecolor{dgray} {rgb}{0.85,0.85,0.85}
\definecolor{vdgray}{rgb}{0.80,0.80,0.80}
\newenvironment{myfont}{\fontfamily{lmtt}\selectfont}{}
  {\color{black}}%
  {}

\begin{document}

\title{Leaky Nets: Recovering Embedded Neural Network Models and Inputs through Simple Power and Timing Side-Channels -- Attacks and Defenses}


\author{\IEEEauthorblockN{
Saurav Maji,
Utsav Banerjee, and 
Anantha P. Chandrakasan}

\IEEEauthorblockA{
Dept. of EECS, Massachusetts Institute of Technology, Cambridge, MA, USA
}

\thanks{Corresponding author: \textit{Saurav Maji} (email: smaji@mit.edu)}
\thanks{\textcopyright $\,$ 2021 IEEE. Personal use of this material is permitted. Permission from IEEE must be obtained for all other uses, in any current or future media, including reprinting/republishing this material for advertising or promotional purposes, creating new collective works, for resale or redistribution to servers or lists, or reuse of any copyrighted component of this work in other works.}
\thanks{A revised version of this paper was published in the IEEE Internet of Things Journal (JIoT) - DOI: \href{https://dx.doi.org/10.1109/JIOT.2021.3061314}{10.1109/JIOT.2021.3061314}}
}

\maketitle

\begin{abstract}
 With the recent advancements in machine learning theory, many commercial embedded micro-processors use neural network models for a variety of signal processing applications. However, their associated side-channel security vulnerabilities pose a major concern. There have been several proof-of-concept attacks demonstrating the extraction of their model parameters and input data. But, many of these attacks involve specific assumptions, have limited applicability, or pose huge overheads to the attacker. In this work, we study the side-channel vulnerabilities of embedded neural network implementations by recovering their parameters using timing-based information leakage and simple power analysis side-channel attacks. We demonstrate our attacks on popular micro-controller platforms over networks of different precisions such as floating point, fixed point, binary networks. We are able to successfully recover not only the model parameters but also the inputs for the above networks. Countermeasures against timing-based attacks are implemented and their overheads are analyzed. 
\end{abstract}

\section{Introduction}
Machine learning (ML), particularly with neural network (NN)-based approaches, have become the de-facto solution for diverse applications such as image recognition \cite{krizhevsky_alexnet_2012}, medical diagnosis \cite{tajbakhsh_cnnbio_2016} and even game theory \cite{david_chess_2016}. Rapid progress in ML theory has led to the deployment of these neural networks on edge devices. Recent years have witnessed a large number of neural network hardware accelerators designed on ASIC as well as FPGA platforms \cite{sze_surveynn_2017}, highlighting the popularity of NNs for achieving energy-efficient inference. Most commercial micro-controllers are equipped with data acquisition units, peripherals, communication interfaces and even radio frequency (RF) modules, making them an extremely attractive choice for implementing system-on-module (SOM) solutions \cite{lu_wearable,qiu_embc,smaji_cicc}. Therefore, many NN algorithms have been implemented on such low-cost micro-controllers \cite{gural2019memory,sparse_nips_2019,heller_embc} for achieving energy-efficient sensing and decision making.  

Embedded neural network implementations, e.g., in healthcare electronics, use locally stored models which have been trained using private data \cite{maier_ctnn} and are considered as intellectual property (IP) of the organizations training them. \cite{shokri_model_2017} had investigated that machine learning models could leak sensitive information about the individual data records over which the model was trained. For many critical applications, like patient-specific diagnosis, the NN model contains private information about the patient and should never be compromised because of privacy concerns. In many of these situations, the NN models provide a competitive edge to the organization or individuals involved, and hence must not be disclosed. Recent years have witnessed new techniques of adversarial attacks on neural network models. These adversarial attacks can sometimes be easier to mount if the underlying NN model is known (known as white-box attacks \cite{chakraborty_adversarial}). All the above discussions strongly motivate the need to keep the neural network model secret.  

Additionally, the inputs to the neural network must also be protected from being recovered by adversaries and eavesdroppers. In all medical applications, the inputs to the neural network are user-specific data that should not be compromised for obvious privacy concerns \cite{nass_medical_2009}. As the raw sensor data is directly fed to the NN, attacking the first NN layer is a preferred choice for recovering the input. 

Side-channel attacks (SCA) \cite{spreitzer_sca_2018} are a major concern in embedded systems where physical access to the device can allow attackers to recover secret data by exploiting information leakage through power consumption, timing, and electromagnetic emanations. 
Common SCA attacks belong to the following two categories \cite{spreitzer_sca_2018}: (a) \textit{Simple power analysis} (SPA) which uses the coarse-grained data dependencies in timing, power consumption or electromagnetic (EM) emanations for identifying the secret value and (b) \textit{Differential power analysis} (DPA) which involves statistical analysis of data collected from ensemble of operations to extract the secret value through fine-grained data dependencies in power consumption or EM emanations. SCA attacks have traditionally been applied for recovering the secret cryptographic keys from the side-channel information \cite{spreitzer_sca_2018,kocher_sca_1996,dhem1998practical,banerjee_aes_2015}. However, side-channel attacks on micro-controller-based NN implementations are also gaining popularity \cite{hua_attack_2018,batina_model_2019,wei_psaip_2018,batina_input_2019,dong_smartiot}.
In this work, we focus on exploiting the side-channel vulnerabilities of embedded NN implementations to recover their model parameters and inputs with simple power / timing analysis and providing optimized countermeasures against timing-based attacks. 
\section{Related Work}

Side-channel analysis for attacking neural network implementations has started to gain importance in the recent years. \cite{hua_attack_2018} reverse engineered two popular convolutional neural networks (AlexNet and SqueezeNet) using memory and timing side-channel leakages from off-chip memory access patterns due to adaptive zero pruning techniques. \cite{batina_model_2019} recovered the complete model of the NN operating on floating point numbers through electromagnetic side-channels using correlation power analysis (CPA) \cite{brier_cpa_2004}, which is a special case of DPA, over the multiplication operations. There have also been few attacks targeting the recovery of network inputs. The inputs of an FPGA-based NN accelerator for MNIST dataset were recovered in \cite{wei_psaip_2018} from the power traces using background model recovery and template matching techniques.  \cite{batina_input_2019} used horizontal power analysis (HPA) \cite{clavier_hca_2010} to predict the input using side-channel leakage from electromagnetic emanations, by correlating the waveform of each multiplication operation with Hamming weight model of the product. \cite{dong_smartiot} used timing side-channel from floating point operations to predict the inputs.

Previous work in this field primarily uses correlation-based attacks \cite{batina_model_2019,batina_input_2019} or power template attacks \cite{wei_psaip_2018} on floating point computations, which involve significant memory and computation overheads in terms of storage and processing of a large number of measured waveforms. \cite{hua_attack_2018} assumes knowledge of the memory access patterns, which may not be always applicable. Furthermore, the accuracy of recovery for correlation-based methods is largely dependent on the signal-to-noise ratio (SNR) of the power waveforms \cite{wei_psaip_2018} or on the number of neurons and size of the model parameters \cite{batina_model_2019,batina_input_2019}, which limits the applicability of these methods only to implementations with high SNR or larger models. For input recovery, \cite{wei_psaip_2018} assumes complete knowledge of the model parameters in the first layer, which is not always possible. The attacks for \cite{wei_psaip_2018,batina_input_2019,dong_smartiot} have been demonstrated only on the MNIST dataset with hand-written digits \cite{mnist_dataset}. Furthermore, the defenses for these attacks have not been implemented and neither have their overheads been analyzed.

\section{Our Contributions}
The key contributions of our work are as follows:
\begin{enumerate}
    \item Our demonstrated attacks involve \textbf{inexpensive timing-based side-channel and simple power analysis (SPA) techniques}. The timing / SPA attacks operate in real-time and are much easier to demonstrate.
    
    \item Our proposed techniques of NN model / input recovery are \textbf{minimally constrained in terms of SNR of measured waveforms and model complexity}.
    
    \item Our techniques have been applied to \textbf{neural networks with different precisions} (e.g. floating point, fixed point, binary NNs) and \textbf{diverse inputs} (MNIST\cite{mnist_dataset}, CIFAR-10 \cite{cifar10_dataset} and ImageNet \cite{imagenet_cvpr09}).
    
    \item Our attacks have been demonstrated over \textbf{multiple embedded micro-processor platforms} such as ATmega-328P, ARM Cortex-M0+ and RISC-V. 
    
    \item \textbf{Software countermeasures} against timing-based attacks have been proposed and their implementation overheads were also analyzed.

\end{enumerate}
\begin{figure}[!ht]
  \centering
  {\includegraphics[width=1\linewidth]{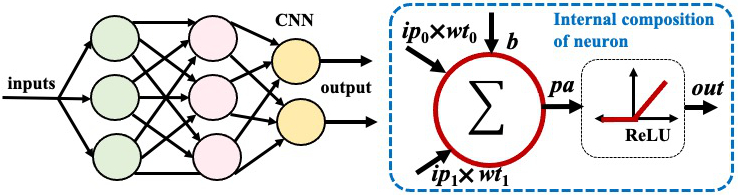}}
  \caption{(left) Organization of a convolutional neural network, and (right) Internal composition of an individual neuron.}
  \label{fig:cnn}
\end{figure}

\section{Threat Model}
\subsection{Attack Scenario}
We will be working with \textit{convolutional neural networks} (CNNs) because of their wide applicability to popular applications \cite{krizhevsky_alexnet_2012,tajbakhsh_cnnbio_2016,sze_surveynn_2017}. A CNN comprises a number of neurons arranged in layers. The  neuron of a given layer receives some input values, does some computations and propagates its output to its successive layer. The neuron multiplies the input $ip$ with corresponding weights $wt$, accumulates them along with the bias $b$ and generates the pre-activation $pa$. Thus, every neuron essentially performs an ensemble of \textit{multiply-accumulate} (MAC) operations ($pa=\sum {ip\times wt} + b$) and then passes $pa$ through a non-linear activation function to generate its output $out$ (Fig. \ref{fig:cnn}). We will use the commonly used non-linear activation function ReLU (\textit{Rectified Linear Unit}) which produces $out = pa$ for $pa \geq 0$ and  $out=0$ for $pa < 0$. For common applications, the final layer outputs the class corresponding to the maximum $pa$ as the classified output category. Hence, we will use this comparison operation as the non-linear activation function for the final layer. 

The scope of this paper is only related to timing-based and SPA-based side-channel techniques. In this paper, we demonstrate that our proposed attacks are easier to perform and the data extraction process is much simpler compared to CPA, DPA and other statistical attacks that require larger amounts of data. However, please note that the statistical attacks are practical (as have been demonstrated in \cite{batina_model_2019,batina_input_2019}) because of the high SNR of micro-controller platforms, and must also be considered for overall side-channel security. We will briefly discuss these attacks in Section \ref{sec:future_work}, where we show that countermeasures against timing attacks may still be susceptible to higher-order statistical attacks.

\subsection{Attacker’s Capabilities}
We consider a passive attacker, whose functionality is very similar to that of an eavesdropper. The following assumptions are considered for the attacker:

\begin{itemize}
\item The attacker is capable of measuring timing and power side-channel information leaked from the implementation of the NN without interrupting the normal execution.  

\item The attacker is capable of recovering the exact sequence of execution of the operations  (e.g. multiply, add, non-linear activation, etc). \cite{batina_model_2019} has shown that these individual operations have distinct EM signatures to identify them. However in this paper, we use the power waveforms to identify the operations being executed (Section \ref{sec:exp_setup}). 

\item \textbf{Model recovery:} The attacker is assumed to have full control over the inputs of the neural network, i.e. the attacker can feed crafted inputs to the network and observe associated power / EM waveforms. However, the adversary cannot change the format / precision of the input. For common image recognition applications, the input is an 8-bit unsigned integer. Thus, the neural network accepts inputs only in this format.

\item \textbf{Input recovery:} The attacker receives side-channel information from execution of the NN. However, no information is assumed to leak from other sources. 
\end{itemize}
We elucidate the threat model using the example of a fitness tracker (Fig. \ref{fig:threat_assumption}) which contains a NN model for classifying ECG signals. The fitness tracker can be separately characterized by providing controlled inputs and observing the side-channel leakages (timing, power and EM side-channels). For recovering the user input, the fitness tracker operates in its normal mode (collecting user's ECG data and classifying it). The timing / power / EM information are collected during this mode using low-power trojans or EM probes and processed to recover the private user data. Fig. \ref{fig:threat_assumption} assumes both power / EM side-channels for a generic use-case. However, we will restrict our study to only the use of power waveform.

\begin{table}[!t]
\scriptsize
\centering
\begin{tabular}{>{\centering}p{0.47\linewidth}>{\centering}p{0.47\linewidth}}
{\multirow{12}{*}{\includegraphics[width=1\linewidth]{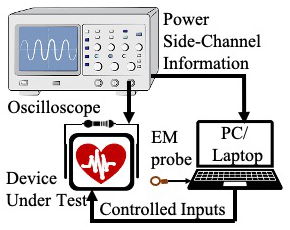}}} &
{\multirow{12}{*}{\includegraphics[width=1\linewidth]{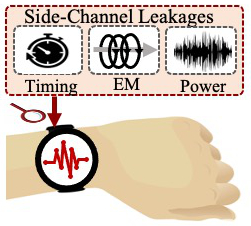}}} \cr
& \cr
& \cr
& \cr
& \cr
& \cr
& \cr
& \cr
& \cr
& \cr
& \cr
& \cr
{\hspace{50mm}\normalsize{(a)}} & {\hspace{50mm}\normalsize{(b)}} \cr
\end{tabular}
\captionof{figure}{Attack scenario for embedded neural network (a) model parameter recovery and (b) input data recovery. }
\label{fig:threat_assumption}
\end{table}
\section{Experimental Setup} \label{sec:exp_setup}
\begin{figure}[!ht]
\centering
\includegraphics[width=0.9\linewidth]{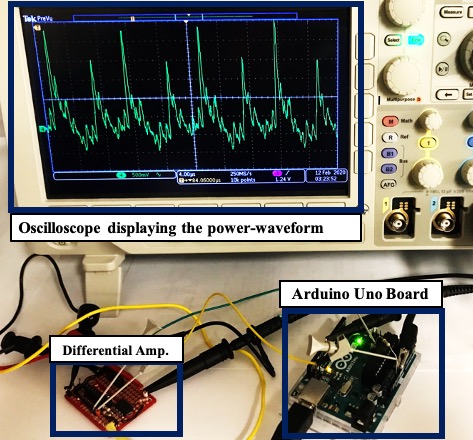}
\caption{Measurement setup for ATmega328P micro-controller, showing the target board with current-sensing differential amplifier and oscilloscope for capturing power waveforms.}
\label{fig:measurement_setup}.
\end{figure}

Our attack methodology and proposed countermeasures are experimentally demonstrated on the widely used ATmega328P, ARM Cortex-M0+ and RISC-V micro-processors. The current consumption is measured by placing a small series resistance ($R_s$ = $10 \Omega$) between the power supply and the supply pin of the chip. The voltage difference across $R_s$ is amplified using an AD8001 current feedback amplifier and the waveform is captured using a high-speed Tektronix MDO3024 Mixed Domain Oscilloscope at a sampling rate of 250 MSamples/s (Fig. \ref{fig:measurement_setup}). 

We demonstrated our attacks on three popular IoT platforms. Atmel ATmega328P \cite{microchip_atmega} and ARM Cortex-M0+ \cite{microchip_arm_cm0plus} micro-processors are very popular commercial platforms for embedded applications and provide high signal-to-noise ratio (SNR) side-channel measurements. RISC-V micro-processors \cite{li_riscv} are gaining popularity for low-power signal processing at IoT edge nodes, so we also evaluate our techniques on a custom-designed RISC-V chip \cite{banerjee_jssc_2019, banerjee_tches_2019}.  

\begin{figure}[!ht]
  \centering
  {\includegraphics[width=1\linewidth]{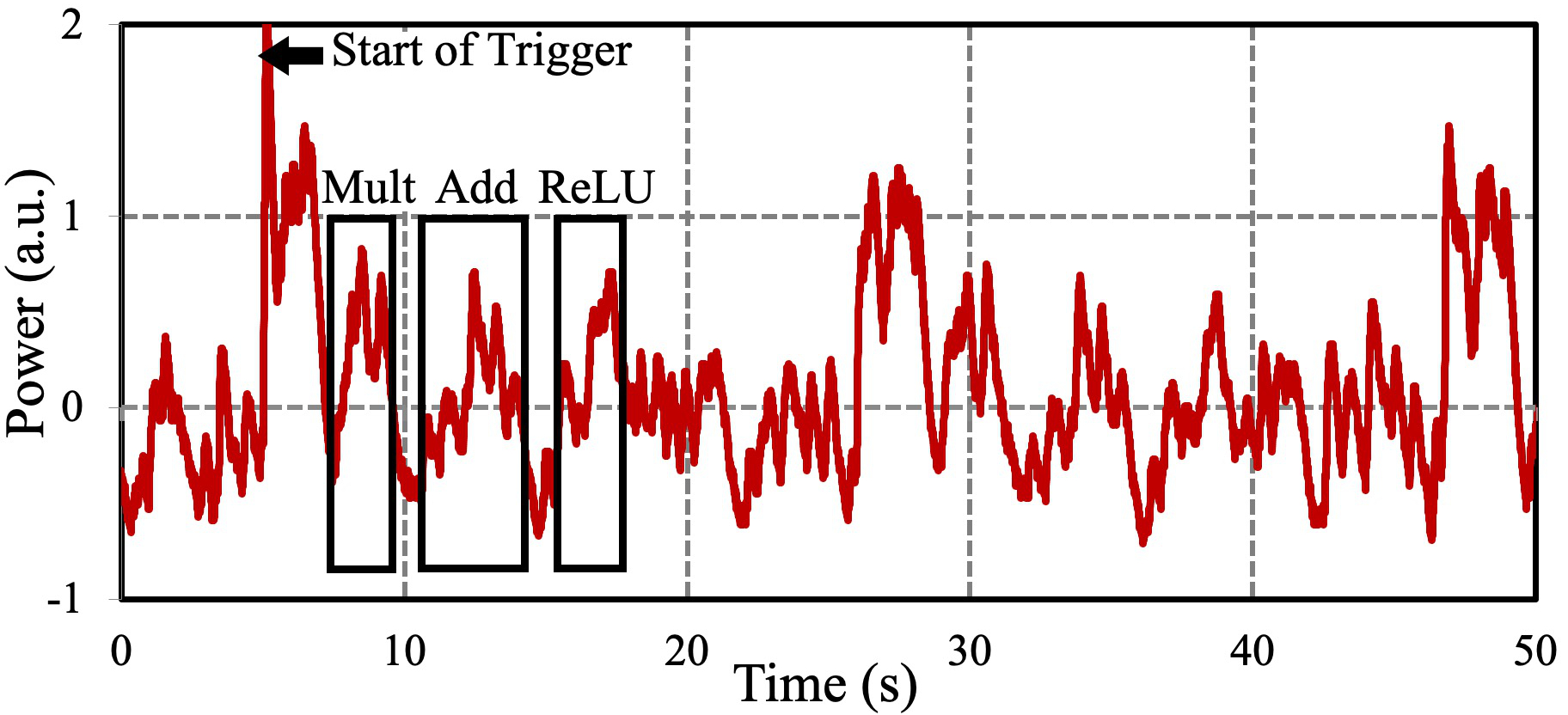}}
  \caption{Identification of various operations from the observed power waveform.}
  \label{fig:op_identify}
\end{figure}

The common step for every attack is the identification of the relevant NN operations (e.g. multiplication, addition, ReLU) from the power waveform. The power waveform consists of distinct power peaks corresponding to these computations. Fig. \ref{fig:op_identify} shows an example of the characteristic power waveform with multiplication, addition, and ReLU operation. A trigger signal is used to automate the data capture process through the oscilloscope. Simple signal processing techniques such as windowing, correlation, and template matching algorithms based on prior characterization can be additionally used to refine this process. \cite{batina_model_2019} and \cite{hua_attack_2018} have described attacks that use coarse-grained parameters to extract the macro-features of the NN model (e.g. number of layers, number of neurons). In this work, we assume that these parameters are extracted using similar methods. We will instead focus on recovering the micro-parameters (e.g., weights, bias, etc).

Commercial micro-controller platforms contain peripheral units like interrupt controllers, serial communication interfaces and data converters which may significantly affect the power consumption. However, we have rigorously ensured throughout this work that none of the peripherals are active during the execution of the NN so as not to affect the identification of the NN operations from power and timing measurements. For better refining the data acquisition process, the measurements of the NN operations are synchronized using appropriate trigger signals to indicate beginning and end of the computation. This assumption models the real-life scenario quite well. In a real use-case, the micro-controller will acquire the data and then execute the NN for appropriate decision making. Therefore, with high probability, all peripheral activities will be observed only before and after the NN operation. In order to further refine the identification process, a combination of both power and EM-based signal acquisition techniques along with improved signal processing techniques can also be used.

In Sections \ref{sec:model_recovery} and \ref{sec:input_recovery}, we first demonstrate all our proposed attacks on the ATmega328P micro-controller \cite{microchip_atmega} available on the Arduino Uno board \cite{arduino_uno}. Then, we extend them to Cortex-M0+ and RISC-V in Section \ref{sec:other_platform}.
\section{Model Recovery}\label{sec:model_recovery}

We will now discuss our proposed methods to extract the model parameters for neural networks with following precisions: floating point (Section \ref{sec:model_flp}), fixed point (Section \ref{sec:model_fxp}), binary NNs (Section \ref{sec:model_bnn}).

\subsection{Floating point neural networks}\label{sec:model_flp}

Floating point NNs operate on real numbers, which are represented according to the IEEE-754 format \cite{2019_ieee754}. This 32-bit number $x=\left(x\textsubscript{31}...x\textsubscript{0}\right)\textsubscript{2}$ is comprised of: 1 sign bit $x\textsubscript{31}$, 8 biased exponent bits $x\textsubscript{30}...x\textsubscript{23}$ and 23 mantissa bits $x\textsubscript{22}...x\textsubscript{0}$. The number $x$ is computed as follows, as shown in eq. \ref{eq:ieee_flp}.
\begin{equation} \label{eq:ieee_flp}
x = (-1)\textsuperscript{\textit{x}\textsubscript{31}} \times 2\textsuperscript{(\textit{x}\textsubscript{30}...\textit{x}\textsubscript{23})\textsubscript{2}-127} \times ({1.x\textsubscript{22}..x\textsubscript{0}})\textsubscript{2} 
\end{equation}

\textbf{Specifications:} We will demonstrate our attack on the example neuron of TABLE \ref{tab:flp_neuron} (with the corresponding weights and bias displayed in its $1^{st}$ column). We will assume that the inputs $ip's$ are constrained to be unsigned 8-bit integers. This is very similar to that of an actual NN where the $1^{st}$-layer neurons accepts the actual data (integer inputs). This is also the worst case scenario in terms of maximally constrained inputs (constrained to be integers).  However, our discussed methodology can be applied for any choice of inputs.

\begin{table}[!h]
\captionof{table}{\footnotesize \MakeTextUppercase{Example Neuron for Floating Point-based Model Recovery (with the displayed weights and bias)}}
\centering
\begin{tabular}{|>{\centering}p{27mm}|>{\centering}p{11mm}|>{\centering}p{15mm}|>{\centering}p{17mm}|}
\hline
\multirow{2}{*}{\scriptsize{PARAMETERS}}    &
\multirow{2}{*}{\scriptsize{MANTISSA}}     &
\multirow{2}{*}{\scriptsize{ZERO}} &
\multirow{2}{*}{\scriptsize{PARAMETERS}}   \cr
\multirow{2}{*}{\tiny{(ACTUAL)}}    &
\multirow{2}{*}{\tiny{(RECOVERED)}}    &
\multirow{2}{*}{\scriptsize{CROSS. IP.}}      &
\multirow{2}{*}{\tiny{(RECOVERED)}}      \cr
& & & \cr
\hline
\cellcolor{white}{$wt\textsubscript{0}=\hspace{2.5mm}1.0390\times2\textsuperscript{-2}$}  & 
\cellcolor{white}{1.0391}  & 
\cellcolor{white}{$ip_{0}^{(1)}=196$} & 
\cellcolor{white}{$\hspace{2mm}1.0391\times2\textsuperscript{0}$}  \cr
\hline
\cellcolor{white}{$wt\textsubscript{1}=-1.6702\times2\textsuperscript{-3}$}  & 
\cellcolor{white}{1.6641}  & 
\cellcolor{lgray}{$ip_{1}^{(2)}=\hspace{1mm}74$}  & 
\cellcolor{lgray}{$-1.6641\times2\textsuperscript{-1}$}  \cr
\hline
\cellcolor{white}{$wt\textsubscript{2}=-1.0855\times2\textsuperscript{-6}$}  & 
\cellcolor{white}{1.0859}  & 
\cellcolor{dgray}{$ip_{2}^{(3)}=213$}  & 
\cellcolor{dgray}{$-1.0859\times2\textsuperscript{-4}$}  \cr
\hline
\cellcolor{white}{$wt\textsubscript{3}=\hspace{2.5mm}1.1803\times2\textsuperscript{-2}$}  & 
\cellcolor{white}{1.1797}  & 
\cellcolor{white}{$ip_{3}^{(1)}=173$}  & 
\cellcolor{white}{$\hspace{2mm}1.1797\times2\textsuperscript{0}$}  \cr
\hline
\cellcolor{white}{$wt\textsubscript{4}=\hspace{2.5mm}1.1255\times2\textsuperscript{-7}$}  & 
\cellcolor{white}{1.1250}  & 
\cellcolor{dgray}{$ip_{4}^{(3)}=188$}  & 
\cellcolor{dgray}{$\hspace{2mm}1.1250\times2\textsuperscript{-5}$}  \cr
\hline
\cellcolor{white}{{$b\hspace{1.5mm}=-1.5906\times2\textsuperscript{5}$}}  & 
\multicolumn{3}{c|}{$-196 \times 1.0391 \times 2\textsuperscript{0} = -1.5911 \times 2\textsuperscript{7}$}  \cr
\hline

\multicolumn{4}{l}{\textbf{\textit{Note:}} The cells corresponding to zero-crossover inputs $ip_{k}^{(1)}$, $ip_{k}^{(2)}$ and}\cr 
\multicolumn{4}{l}{ $ip_{k}^{(3)}$ along with its associated recovered weights are marked in white,} \cr

\multicolumn{4}{l}{light gray and dark gray colors respectively.} \cr
\end{tabular}
\label{tab:flp_neuron}
\end{table}

\textbf{Methodology:} Steps I-IV describe the detailed methods for extracting the parameters of the $1\textsuperscript{st}$ layer, while step V shows how to extend it to successive layers.

\textbf{Step I. Extraction of mantissas of weights ($1.m\textsubscript{wt}$):}
The timing side channel information from floating point-based multiplication operation is used to recover the weight mantissa $1.m\textsubscript{wt}$. Let $T(ip\times{wt})$ denote the time (in cycles) taken to perform the multiplication of weight $wt$ and input activation $ip$. From our characterization of the floating point multiplications on the ATmega328P platform, we found that the timing of the multiplication operation $T(ip\times{wt})$ is dependent on the mantissa of the operands ($1.m\textsubscript{ip}$ and $1.m\textsubscript{wt}$) and is independent of the exponents involved. Hence, we can denote this mathematically as $T(ip\times{wt}) \equiv T(1.m\textsubscript{ip}\times1.m\textsubscript{wt})$.

\begin{figure}[!ht]
  \centering
  {\includegraphics[width=0.90\linewidth]{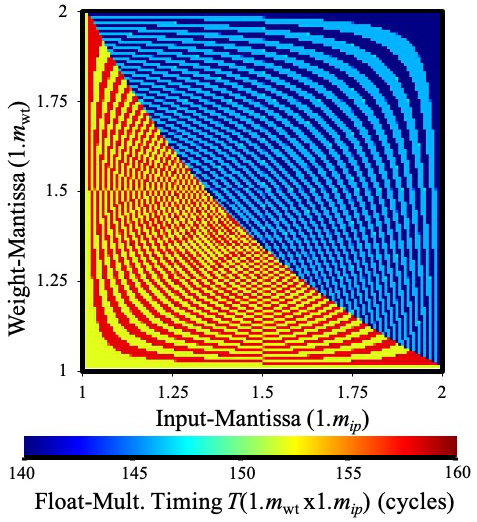}}
  \caption{Timing of $T(1.m\textsubscript{ip}\times 1.m\textsubscript{wt})$ for different weight mantissas $1.m\textsubscript{wt}$ (eq. \ref{eq:flp_wtm}) and input mantissas $1.m\textsubscript{ip}$ (eq. \ref{eq:flp_ipm})}
  \label{fig:flp_mantissa_lut}
\end{figure}

For a given weight mantissa $1.m\textsubscript{wt}$, we obtain a mapping between all possible input activation mantissas $1.m\textsubscript{ip}$ $\left( \text{i.e.,~} 1.m\textsubscript{ip}\in[1,2)\right)$ and the timing of the corresponding multiplication operation $T(1.m\textsubscript{ip}\times1.m\textsubscript{wt})$. This mapping is denoted as  $1.m\textsubscript{ip}\xrightarrow[1.m\textsubscript{ip}\in[1,2)]{1.m\textsubscript{wt}}T(1.m\textsubscript{ip}\times{1.m\textsubscript{wt}})$. Fig. \ref{fig:flp_mantissa_lut} shows the $T(1.m\textsubscript{ip} \times 1.m\textsubscript{wt})$ with the horizontal axis displaying $1.m\textsubscript{ip}$ and the vertical axis displaying $1.m\textsubscript{wt}$. Each column represents the mapping $1.m\textsubscript{ip}\xrightarrow[1.m\textsubscript{ip}\in[1,2)]{1.m\textsubscript{wt}}T(1.m\textsubscript{ip}\times{1.m\textsubscript{wt}})$ corresponding to its weight mantissa $1.m\textsubscript{wt}$. Instead of exhaustive range of $1.m\textsubscript{ip}\in[1,2)$, we only consider the following input mantissas of  eq. \ref{eq:flp_ipm} for constructing this look-up table (LUT).

\begin{equation}\label{eq:flp_ipm}
1.m\textsubscript{ip} = 1\frac{ip'}{128},ip'\in\{0,1,...,127\} 
\end{equation}

As discussed earlier, the mantissa contains 23 bits. However, as shown in \cite{batina_model_2019}, it becomes sufficient to extract only the 7 most significant bits of the mantissa. The error introduced because of this truncation is less than 1\%. Also, floating point numbers with 7 mantissa bits (instead of 23 mantissa bits), and usual 7 exponent bits and 1 sign bit, (known as \texttt{bfloat16} \cite{kalamkar_bfloat_2019}) are becoming popular. Thus, only the following weight mantissas are considered in the LUT of Fig. \ref{fig:flp_mantissa_lut}:
\begin{equation}\label{eq:flp_wtm}
1.m\textsubscript{wt} = 1\frac{wt'}{128}, wt' \in \{0,1,...,127\} 
\end{equation}

From our characterization of the floating point-based multiplication operations on the Arduino ATmega328P platform, we find that $1.m\textsubscript{ip}\xrightarrow[1.m\textsubscript{ip}\in[1,2)]{1.m\textsubscript{wt}}T(1.m\textsubscript{ip}\times{1.m\textsubscript{wt}})$ is unique for every $1.m\textsubscript{wt}$. In other words, no two columns of Fig. \ref{fig:flp_mantissa_lut} are same. An unknown weight mantissa $1.m\textsubscript{wt\textsubscript{0}}$ is obtained by correlating its $1.m\textsubscript{ip} \xrightarrow[1.m\textsubscript{ip}\in[1,2)]{1.m\textsubscript{wt\textsubscript{0}}} T(1.m\textsubscript{ip} \times 1.m\textsubscript{wt\textsubscript{0}})$ mapping with all the columns of Fig. \ref{fig:flp_mantissa_lut} and selecting $1.m\textsubscript{wt}$ corresponding to the highest correlated column. The actual and the extracted mantissas for the example neuron are shown in TABLE \ref{tab:flp_neuron}, and they match very closely.

\begin{figure}[!ht]
  \centering
  {\includegraphics[width=0.90\linewidth]{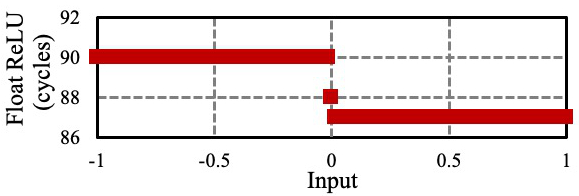}}
  \caption{Timing side-channel for floating point ReLU.}
  \label{fig:flp_relu}
\end{figure}

\textbf{Step II. 1$^{st}$ round of weight and bias extraction:} For the following steps, we utilize the timing side-channel of the ReLU operation. As shown in Fig. \ref{fig:flp_relu}, the timing of the ReLU operation is  dependent on the inputs. For $+$ve inputs, the execution of the ReLU operation takes fewer cycles, whereas for $-$ve inputs, it uses more cycles.

For determining the exponents, we utilize the concept of \textbf{\textit{zero-crossover input}} (very similar to the concept described in \cite{hua_attack_2018}). We define the zero-crossover input $ip\stackanchor{\scriptsize{(1)}}{\scriptsize{k}}$ as the valid input of $ip\textsubscript{k}$ for which the following  equation $\left( ip\textsubscript{k}\times wt\textsubscript{k} + b \right)$ crosses $0$. This means that  $\left( \left(ip\stackanchor{\scriptsize{(1)}}{\scriptsize{k}}-1\right)\times wt\textsubscript{k} + b \right)$ and $\left(ip\stackanchor{\scriptsize{(1)}}{\scriptsize{k}}\times wt\textsubscript{k} + b \right)$ are of opposite signs. To obtain the value of the zero-crossover input $ip\stackanchor{\scriptsize{(1)}}{\scriptsize{k}}$, all the other inputs except $ip\textsubscript{k}$ of the neuron are set to $0$. This forces the pre-activation function to be $pa = \left(ip\textsubscript{k}\times wt\textsubscript{k} + b \right)$. Now, $ip\textsubscript{k}$ is incremented successively from $0$ to $255$ and the timing of the ReLU operation is observed. The distinct timings of the ReLU operation for $+$/$-$ inputs can be used to detect when $pa$ changes its sign. The value of the input for which this transition occurs is the zero-crossover input $ip\stackanchor{\scriptsize{(1)}}{\scriptsize{k}}$. We determine the value of this zero-crossover input corresponding to every input / weight pair. However, not all input / weight pair will have a zero-crossover input. The zero-crossover input $ip\stackanchor{\scriptsize{(1)}}{\scriptsize{k}}$ is obtained only when $b$ and $wt\textsubscript{k}$ are of opposite signs and $\lvert wt\textsubscript{k}\times255\rvert > \lvert b\rvert$. As shown in eq. \ref{eq_flp_crossover1}, at zero-crossover input $ip\stackanchor{\scriptsize{(1)}}{\scriptsize{k}}$,  $\left(ip\stackanchor{\scriptsize{(1)}}{\scriptsize{k}}\times wt\textsubscript{k} + b \right)\approx0$, and hence $wt\textsubscript{k}$ can be expressed in terms of $b$ and $ip\stackanchor{\scriptsize{(1)}}{\scriptsize{k}}$. 

\begin{equation} \label{eq_flp_crossover1} 
    \left(ip\stackanchor{\scriptsize{(1)}}{\scriptsize{k}} \times wt\textsubscript{k} +b\right)\approx0 \Rightarrow ip\stackanchor{\scriptsize{(1)}}{\scriptsize{k}}\approx
    \frac{-b}{wt\textsubscript{k}} 
\end{equation}

TABLE \ref{tab:flp_neuron} shows the obtained $ip\stackanchor{\scriptsize{(1)}}{\scriptsize{k}}$ for our example neuron. For many weights (e.g., $wt\textsubscript{1}, wt\textsubscript{2}, \text{and } wt\textsubscript{4}$), $ip\stackanchor{\scriptsize{(1)}}{\scriptsize{k}}$  could not be obtained because either $b$ and $wt\textsubscript{k}$ are of same sign or because $\lvert wt\textsubscript{k}\times255\rvert < \lvert b\rvert$. For these weights $wt\textsubscript{k}$, the sign of the ReLU function remains unchanged for $ip\textsubscript{k}\in\{0,...,255\}$

We will now exploit the obtained values of the zero-crossover inputs $ip\stackanchor{\scriptsize{(1)}}{\scriptsize{k}}$ to determine the value of the corresponding weights $wt\textsubscript{k}$. We can determine the exponents of the weights correct only up to a constant factor. Let us select any weight $wt\textsubscript{ref}$ with a valid $ip\stackanchor{\scriptsize{(1)}}{\scriptsize{ref}}$ value as the reference weight (e.g. for our neuron, we choose $wt\textsubscript{0}$ as the reference weight) and let us denote its exponent to be $e\textsubscript{ref}$, that is unknown. We would now determine the exponents of all the other weights relative to $e\textsubscript{ref}$. Thus, all the weights would be computed correct to the unknown factor $2\textsuperscript{e\textsubscript{ref}}$. The exponent $e\textsubscript{k}$ of the weight $wt\textsubscript{k}$ is obtained with respect to $e\textsubscript{ref}$ using eq. \ref{eq_flp_exponent1}.      
\begin{equation} \label{eq_flp_exponent1}
e\textsubscript{k}-e\textsubscript{ref}=\left\lceil \text{log}\textsubscript{2}\left(\frac{1.m\textsubscript{ref}}{1.m\textsubscript{k}} \times \frac{ip\stackanchor{\scriptsize{(1)}}{\scriptsize{ref}}}{ip\stackanchor{\scriptsize{(1)}}{\scriptsize{k}}}\right)\right\rfloor
\end{equation}

The $\lceil{x}\rfloor$ operation of eq. \ref{eq_flp_exponent1} rounds $x$ to its nearest integer. Please refer to Appendix \ref{app:flp_wt} for the derivation of the equations. The values of $e\textsubscript{k} - e\textsubscript{ref}$ obtained using eq. \ref{eq_flp_exponent1} is shown in TABLE \ref{tab:flp_neuron}. As shown in TABLE \ref{tab:flp_neuron}, the estimated exponents are correct to a constant difference of $-2$. This factor of $e\textsubscript{ref}=-2$ remains unknown to the attacker. We will show in Step V that the unknown scaling factor does not affect the correctness of the computation.

The value of sgn($b$) is obtained by setting all the inputs ${ip\textsubscript{k}}=0$ (thus, ensuring that $pa = b$) and then observing the timing of the ReLU operation. 
The value of sgn($wt\textsubscript{k}$) for all those indexes (${k}$) having valid  $ip\stackanchor{\scriptsize{(1)}}{\scriptsize{k}}$ values is opposite to that of sign($b$). For our example neuron, sgn($b$)$=-1$ and hence, sgn($wt\textsubscript{k}$)=1 for all indexes (${k}$) having valid $ip\stackanchor{\scriptsize{(1)}}{\scriptsize{k}}$ values.

The bias $b$ is obtained from eq. \ref{eq_flp_crossover1} (by using $1.m\textsubscript{ref}$ instead of generic weight $1.m\textsubscript{k}$) as shown in eq. \ref{eq_flp_bias1}. This bias $b$ is calculated in TABLE \ref{tab:flp_neuron}.
\begin{equation} \label{eq_flp_bias1}
b \approx -wt\textsubscript{ref}\times ip\stackanchor{\scriptsize{(1)}}{\scriptsize{ref}} =(-1)\textsuperscript{\text{sgn}(b)}\times 1.m\textsubscript{ref}\times2\textsuperscript{e\textsubscript{ref}} \times ip\stackanchor{\scriptsize{(1)}}{\scriptsize{ref}}
\end{equation}

\textbf{Step III. 2$^{nd}$ round of weight extraction:} In order to determine some of the remaining weights, we will use a variant of the zero-crossover input $ip\stackanchor{\scriptsize{(2)}}{\scriptsize{k}}$ as the valid input of $ip\textsubscript{k}$ for which $\left( ip\textsubscript{k}\times wt\textsubscript{k} + 255 \times wt\textsubscript{ref} + b \right)$ crosses $0$.  The choice of $wt\textsubscript{ref}$ remains the same as that of Step II. In order to determine $ip\stackanchor{\scriptsize{(2)}}{\scriptsize{k}}$, we initialize  $ip\textsubscript{ref}$ to $255$ and all the other inputs except $ip\textsubscript{k}$ to $0$. We now increment $ip\textsubscript{k}$ from $0$ to $255$ and observe the change in sign of the ReLU function to detect $ip\stackanchor{\scriptsize{(2)}}{\scriptsize{k}}$.

Extending the discussions from Step II, we find that $ip\stackanchor{\scriptsize{(2)}}{\scriptsize{k}}$ is obtained only when sgn($b$)=sgn($wt\textsubscript{k}$) and $\lvert wt\textsubscript{k}\times255\rvert > \lvert \left(b+255\times wt\textsubscript{ref}\right)\rvert$. Following the analysis in 
Appendix \ref{app:flp_wt}, we can obtain the unknown exponent $e\textsubscript{k}$- $e\textsubscript{ref}$ using eq. \ref{eq_flp_exponent2}.

\begin{equation} \label{eq_flp_exponent2}
e\textsubscript{k}-e\textsubscript{ref}=\left\lceil \text{log}\textsubscript{2}\left(\frac{1.m\textsubscript{ref}}{1.m\textsubscript{k}} \times \frac{\left(255-ip\stackanchor{\scriptsize{(1)}}{\scriptsize{ref}}\right)}{ip\stackanchor{\scriptsize{(2)}}{\scriptsize{k}}}\right)\right\rfloor
\end{equation}

Continuing with $wt\textsubscript{ref}=wt\textsubscript{0}$, TABLE \ref{tab:flp_neuron} shows the obtained $ip\stackanchor{\scriptsize{(2)}}{\scriptsize{k}}$ terms (marked in light-gray color) and the recovered weights for our example neuron.

\textbf{Step IV. 3$^{rd}$ round of weight extraction:} In the previous steps, we were able to recover all the weights, for which $wt\textsubscript{k}$ lies outside the range ${[-\frac{b}{255}-wt\textsubscript{ref},-\frac{b}{255}]}$.
In order to determine the remaining weights $wt\textsubscript{k}$, we define a variant of the zero-crossover input $ip\stackanchor{\scriptsize{(3)}}{\scriptsize{k}}$ as the valid input of $ip\textsubscript{ref}$ for which the following  equation $\left( ip\textsubscript{ref}\times wt\textsubscript{ref} + 255 \times wt\textsubscript{k} + b \right)$ crosses $0$. It should be noted here that index $ip\stackanchor{\scriptsize{(3)}}{\scriptsize{k}}$ corresponds to the input $ip\textsubscript{ref}$ instead of $ip\textsubscript{k}$, i.e. for $ip\stackanchor{\scriptsize{(3)}}{\scriptsize{k}}$, we have the equation $\left(ip\stackanchor{\scriptsize{(3)}}{\scriptsize{k}}\times wt\textsubscript{ref} + 255 \times wt\textsubscript{k} + b \right)$ changing sign. The choice of $wt\textsubscript{ref}$ remains the same as that of Step II.  Following the analysis in 
Appendix \ref{app:flp_wt}, we can obtain the unknown sign and exponent $e\textsubscript{k} - e\textsubscript{ref}$ informations using eq. \ref{eq_flp_sign3} and eq. \ref{eq_flp_exponent3} respectively.

\begin{equation} \label{eq_flp_sign3}
 \text{sgn}(wt\textsubscript{k})=\text{sgn}\left(\frac{ip\stackanchor{\scriptsize{(3)}}{\scriptsize{k}}}{ip\stackanchor{\scriptsize{(1)}}{\scriptsize{ref}}}-1\right) \times \text{sgn}(b)
\end{equation}

\begin{equation} \label{eq_flp_exponent3}
e\textsubscript{k}-e\textsubscript{ref}=\left\lceil \text{log}\textsubscript{2}\left(\frac{1.m\textsubscript{ref}}{1.m\textsubscript{k}} \times \frac{\left\lvert ip\stackanchor{\scriptsize{(3)}}{\scriptsize{k}}-ip\stackanchor{\scriptsize{(1)}}{\scriptsize{ref}}\right\rvert}{255}\right)\right\rfloor
\end{equation}
TABLE \ref{tab:flp_neuron} shows the recovered weights relative to $2\textsuperscript{e\textsubscript{ref}}$ obtained using eq. \ref{eq_flp_sign3} and eq. \ref{eq_flp_exponent3}. For $wt\textsubscript{k} \in {(-\frac{b}{255}-wt\textsubscript{ref},-\frac{b}{255})}$, it is guaranteed that $ip\stackanchor{\scriptsize{(3)}}{\scriptsize{k}} \in \{1,2,...,255\}$. Thus, all the weights can be determined after this step. 

\textbf{Step V. Model recovery from successive layers:} Having determined the weights and biases of the first layer, we now successively determine weights and biases of the next layers. For detecting weight mantissas of the successive layers, we construct suitable lookup tables as per the output of the previous layer. The scaling factors for each layer will get accumulated as we reverse engineer each layer. For a feed-forward neural network, these scaling factors also do not affect the output of the comparison operation to determine the most-probable class. Hence, we can safely ignore the scaling factors while retaining the exact functionality of the original NN.

The above analysis assumes ReLU  operation as the non-linear activation function. This is the worst-case situation in terms of extraction of the weights. Use of other non-linear activation functions like $sigmoid$, $tanh$, $softmax$ or $argmax$ (comparison for finding the maximum class)  leaks more information from its timing side channel (as shown in Fig. \ref{fig:flp_nonlinear_act}) and hence, can be used to easily recover the unknown scaling factors.

\begin{table}[!h]
\scriptsize
\centering
\begin{tabular}{p{0.45\linewidth}p{0.45\linewidth}}
\multicolumn{1}{l}{\multirow{10}{*}{\includegraphics[width=0.45\linewidth]{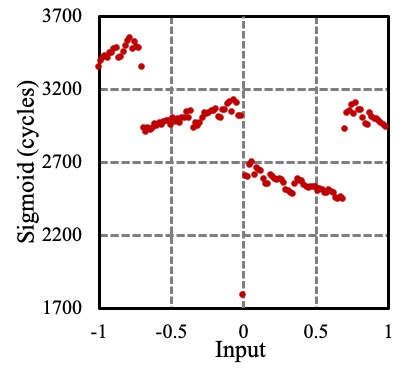}}} & 
\multicolumn{1}{l}{\multirow{10}{*}{\includegraphics[width=0.45\linewidth]{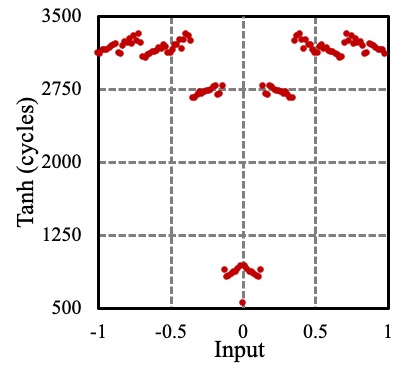}}} \cr
{} & {} \cr
{} & {} \cr
{} & {} \cr
{} & {} \cr
{} & {} \cr
{} & {} \cr
{} & {} \cr
{} & {} \cr
{} & {} \cr
{} & {} \cr
{} & {} \cr
{} & {} \cr
{\hspace{6mm}\fontsize{9}{13}\selectfont $\text{(a)}     \hspace{2mm}sigmoid(x)=\frac{1}{1+e\textsuperscript{-\textit{x}}}$} &
{\hspace{8mm}\fontsize{9}{13}\selectfont $\text{(b)}     \hspace{2mm}tanh(x)=\frac{e\textsuperscript{\textit{x}}-e\textsuperscript{-\textit{x}}}{e\textsuperscript{\textit{x}}+e\textsuperscript{-\textit{x}}}$} \cr
\end{tabular}
\captionof{figure}{Timing side-channel for (a) $sigmoid$ and (b) $tanh$ floating point non-linear activation functions.} 
\label{fig:flp_nonlinear_act}
\end{table}

\subsection{Fixed point neural networks} \label{sec:model_fxp}

Fixed point NNs are extremely popular from the perspective of hardware implementations because their corresponding operations can be mapped very effectively to the processor's arithmetic hardware \cite{sze_surveynn_2017}.

\textbf{Assumptions:} For our example neuron of TABLE \ref{tab:fxp_neuron}, the inputs $ip$'s are 8-bit unsigned integers, whereas the weights ($wt$'s) / bias ($b$'s) are quantized to 4-bit / 8-bit signed integers. The pre-activations $pa$ are rectified using the ReLU operation, quantized to 8-bit unsigned integer and then used as the input to the neuron for the next layer. The bit-precision of the inputs, weights and biases have been chosen here for ease of demonstration. 

\begin{figure}[!ht]
  \centering
  {\includegraphics[width=1.00\linewidth]{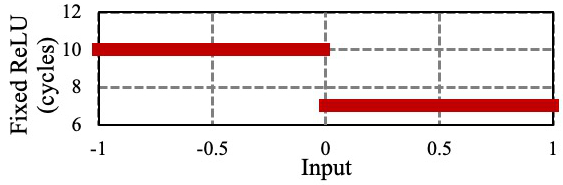}}
  \caption{Timing side-channel for fixed point ReLU computation.}
  \label{fig:fxp_relu}
\end{figure}

\textbf{Methodology:} The steps for recovering fixed point model parameters are described below:

\textbf{Step I. Construction of the lookup table:} The fixed point ReLU, similar to floating point ReLU, suffers from timing side-channel leakage depending on the sign of the operands and hence, can be used to determine the sign of $pa$ (Fig. \ref{fig:fxp_relu}).

\begin{figure}[!ht]
  \centering
  {\includegraphics[width=1.00\linewidth]{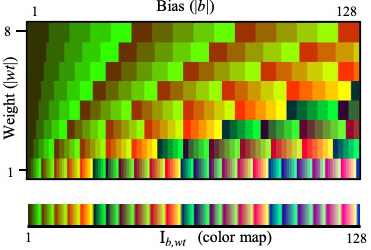}}
  \caption{LUT of $I\textsubscript{b,wt} = \left\lceil\frac{-b}{wt}\right\rceil$ for fixed point model recovery.} 
  \label{fig:fxp_lut}
\end{figure}

We construct a lookup table (LUT) of the zero-crossover inputs $I\textsubscript{\scriptsize{b,wt}} = \left\lceil\frac{\normalsize{-b}}{\normalsize{wt}}\right\rceil$ for every possible weight ($wt$)-bias ($b$) pair. Fig. \ref{fig:fxp_lut} displays this LUT of $I\textsubscript{\scriptsize{b,wt}}$ as a color map, with the x-axis / y-axis displaying the bias / weights respectively. $I\textsubscript{\scriptsize{b,wt}}$ can be determined only when $\text{sgn}(wt)\neq \text{sgn}(b)$.
Without loss of generality, we only consider $wt > 0$ and $b < 0$. Accordingly, the x-axis comprises of $|b| \in \{1,2,..128\}$ and the y-axis comprises of $|wt| \in \{1,2,...,8\}$. It should be noted that this LUT is constructed independent of the targeted neuron and needs to be constructed only once.

\begin{table}[!ht]
\scriptsize
\captionof{table}{\footnotesize \MakeTextUppercase{Example Neuron for Fixed point-based Model Recovery (with the displayed weights and bias) }}
\centering
\begin{tabular}{|p{15mm}|>{\centering}p{17mm}|p{15mm}|>{\centering}p{17mm}|}
\hhline{|-|-|-|-|}
\multirow{3}{*}{\scriptsize{PARAMETERS}} &
{\scriptsize{ZERO-}} &
\multirow{3}{*}{\scriptsize{PARAMETERS}} &
{\scriptsize{ZERO-}} \cr

& {\scriptsize{CROSSOVER}} & 
& {\scriptsize{CROSSOVER}} \cr
& {\scriptsize{INFORMATION}} & 
& {\scriptsize{INFORMATION}} \cr
\hhline{|-|-|-|-|}
\cellcolor{white}&\cellcolor{white}& 
\cellcolor{lgray}&\cellcolor{lgray} \cr
\cellcolor{white}\multirow{-2}{*}
{\normalsize{$wt\textsubscript{0} = -1$}}     & 
\cellcolor{white}\multirow{-2}{*}
{\normalsize{$ip\stackanchor{\scriptsize{(1)}}{\scriptsize{0}} = 108$}}&
\cellcolor{lgray}\multirow{-2}{*}
{\normalsize{$wt\textsubscript{5} =~~2$}}     & 
\cellcolor{lgray}\multirow{-2}{*}
{\normalsize{$ip\stackanchor{\scriptsize{(2)}}{\scriptsize{5}} = 46$}}  \cr
\hhline{|-|-|-|-|}
\cellcolor{white} & \cellcolor{white} & 
\cellcolor{white} & \cellcolor{white} \cr
\cellcolor{white}\multirow{-2}{*}
{\normalsize{$wt\textsubscript{1} = -3$}}     & 
\cellcolor{white}\multirow{-2}{*}
{\normalsize{$ip\stackanchor{\scriptsize{(1)}}{\scriptsize{1}} = 36$}}& 
\cellcolor{white}\multirow{-2}{*}
{\normalsize{$wt\textsubscript{6} = -6$}}     & 
\cellcolor{white}\multirow{-2}{*}
{\normalsize{$ip\stackanchor{\scriptsize{(1)}}{\scriptsize{6}} = 18$}} \cr
\hhline{|-|-|-|-|}
\cellcolor{lgray}&\cellcolor{lgray} & 
\cellcolor{lgray}&\cellcolor{lgray} \cr
\cellcolor{lgray}\multirow{-2}{*}
{\normalsize{$wt\textsubscript{2} =~~4$}}     & 
\cellcolor{lgray}\multirow{-2}{*}
{\normalsize{$ip\stackanchor{\scriptsize{(2)}}{\scriptsize{2}} = 23$}}   &
\cellcolor{lgray}\multirow{-2}{*}
{\normalsize{$wt\textsubscript{7} =~~5$}}     & 
\cellcolor{lgray}\multirow{-2}{*}
{\normalsize{$ip\stackanchor{\scriptsize{(2)}}{\scriptsize{7}} = 19$}}  \cr
\hhline{|-|-|-|-|}
\cellcolor{white} &\cellcolor{white} & 
\cellcolor{white} &\cellcolor{white} \cr
\cellcolor{white}\multirow{-2}{*}
{\normalsize{$wt\textsubscript{3} = -7$}}     & 
\cellcolor{white}\multirow{-2}{*}
{\normalsize{$ip\stackanchor{\scriptsize{(1)}}{\scriptsize{3}} = 16$}}  & 
\cellcolor{white}\multirow{-2}{*}
{\normalsize{$wt\textsubscript{8} =~~0$}}     & 
\cellcolor{white}\multirow{-2}{*}
{\normalsize{$-$}}  \cr
\hhline{|-|-|-|-|}
\cellcolor{white} &\cellcolor{white} & 
\cellcolor{white} &\cellcolor{white} \cr
\cellcolor{white}\multirow{-2}{*}
{\normalsize{$wt\textsubscript{4} = -8$}}     & 
\cellcolor{white}\multirow{-2}{*}
{\normalsize{$ip\stackanchor{\scriptsize{(1)}}{\scriptsize{4}} = 14$}}  & %
\cellcolor{white}\multirow{-2}{*}
{\normalsize{$b =~~108$}}     & 
\cellcolor{white}\multirow{-2}{*}
{\normalsize{}}                                \cr
\hhline{|-|-|-|-|}
\multicolumn{4}{l}{\textbf{\textit{Note:}} The cells corresponding to zero-crossover inputs $ip_{k}^{(1)}$ and $ip_{k}^{(2)}$ are}\cr 
\multicolumn{4}{l}{marked in white and light gray colors respectively.} \cr
\end{tabular}
\label{tab:fxp_neuron}
\end{table} 

\textbf{Step II. Recovery of the bias:} We determine the zero-crossover input $ip\stackanchor{\scriptsize{(1)}}{\scriptsize{k}}$ (defined as the valid value of $ip\textsubscript{k}$ for which  $ip\textsubscript{k}\times wt\textsubscript{k} + b$ crosses $0$), for all the possible  $wt\textsubscript{k}-b$ pairs of the neuron. Similar to Section \ref{sec:model_flp}, $ip\stackanchor{\scriptsize{(1)}}{\scriptsize{k}}$ is obtained by setting all the inputs except $ip\textsubscript{k}$ to 0 and incrementing $ip\textsubscript{k}$ until the ReLU operation changes its sign (observed from its timing side channel). TABLE \ref{tab:fxp_neuron} displays the obtained $ip\stackanchor{\scriptsize{(1)}}{\scriptsize{k}}$ values.

From the ensemble collection of $ip\stackanchor{\scriptsize{(1)}}{\scriptsize{k}}$, we determine the bias $|b|$ by locating the column of the LUT in Fig. \ref{fig:fxp_lut} which uniquely contains all the obtained  $ip\stackanchor{\scriptsize{(1)}}{\scriptsize{k}}$ values. TABLE \ref{tab:fxp_neuron} shows the obtained $ip\stackanchor{\scriptsize{(1)}}{\scriptsize{k}}$ zero-crossover inputs for our example neuron. $|b|=108$ of LUT of Fig. \ref{fig:fxp_lut} uniquely contains all of the obtained $ip\stackanchor{\scriptsize{(1)}}{\scriptsize{k}}$ values. By observing the timing of the ReLU operations after setting all the inputs to 0, we found that the bias is positive and hence, $b=108$.

\textbf{Step III. 1$\textsuperscript{st}$ round of weight recovery:} After determining $b$, we back-calculate weight $wt\textsubscript{k}$ from $ip\stackanchor{\scriptsize{(1)}}{\scriptsize{k}}$ by looking at the column of the LUT corresponding to $b$ and finding the weight corresponding to  $I\textsubscript{b,wt} = ip\stackanchor{\scriptsize{(1)}}{\scriptsize{k}}$. The sign of $wt\textsubscript{k}$ is opposite to that of the sign of bias $b$. In this step, we were able to recover the weights $wt\textsubscript{k}$ for which  sgn($wt\textsubscript{k}$) $\neq$ sgn($b$).

 \textbf{Step IV. 2$\textsuperscript{nd}$ round of weight recovery:} To recover the remaining weights $wt\textsubscript{k}$ with sgn($wt\textsubscript{k}$) = sgn($b$), we select a known weight $wt\textsubscript{ref}$ and fix its corresponding input $ip\textsubscript{ref}$ such that $\left(ip\textsubscript{ref}\times wt\textsubscript{ref}+b\right)$ is also 8-bit signed integer but with opposite sign of $b$. Similar to Step III in Section \ref{sec:model_fxp}, we now obtain the zero-crossover input $ip\stackanchor{\scriptsize{(2)}}{\scriptsize{k}}$ as the valid value of $ip\textsubscript{k}$ for which $\left( ip\textsubscript{k}\times wt\textsubscript{k} + ip\textsubscript{ref} \times wt\textsubscript{ref} + b \right)$ crosses $0$.

For our example, we choose $wt\textsubscript{ref}=wt\textsubscript{0}=-1$ and $ip\textsubscript{ref}=200$ and then obtained  all possible values of $ip\stackanchor{\scriptsize{(2)}}{\scriptsize{k}}$ in light-gray color in TABLE \ref{tab:fxp_neuron}. From the LUT, we were able to determine $wt\textsubscript{k}$ from the LUT column corresponding to $\left|108 + 200\times(-1)\right| = |92|$. For our example neuron, we were able to correctly recover the remaining weights $wt\textsubscript{k}$ for which  sgn($wt\textsubscript{k}$) $=$ sgn($b$). The above steps can then be applied sequentially to determine the parameters of the successive layers in layer-wise fashion. 

It is important to note that for many cases, the weights / bias cannot be uniquely determined. In such situations, the attacker need to suitably determine different variants of zero-crossover inputs under different conditions $\left(\text{e.g.} \left\lceil\frac{-(b+ip\textsubscript{m}\times wt\textsubscript{m})}{wt\textsubscript{k}}\right\rceil \text{or} \left\lceil\frac{-b}{wt\textsubscript{k}+wt\textsubscript{m}}\right\rceil \right)$.

It is important to reiterate that the LUT construction is a one-time process (independent of the NN model to be used) and the same LUT can be used to recover NN parameters as many times as required, thus making our methods extremely practical. The LUT for floating point NN (shown in Fig. \ref{fig:flp_mantissa_lut}) is specific to the underlying micro-controller platform (its variants for ARM Cortex-M0+ and RISC-V RV32IM are shown in Fig. \ref{fig:other_platform}:I(a) and II(a) respectively), whereas the LUT for fixed point NN (shown in Fig. \ref{fig:fxp_lut}) is platform-independent. 
 

\subsection{Binary neural networks} \label{sec:model_bnn}

For binary neural networks (BNNs), the weights $wt$ are constrained to $\pm1$. However, the bias $b$ is a signed integer. Binarized neural networks \cite{2016_courbariaux} are a special class of BNNs where even the activations are also quantized to $\pm1$ (except the inputs to the first layer). BNNs are popular for energy-constrained devices because of reduced memory requirements and elimination of the multiplication operation \cite{nurvitadhi_bnn}. In this section, we will discuss the extraction of the model parameters of binary neural networks. The case of binarized neural network can be dealt with appropriate modifications. 

 \begin{table}[!htp]
\scriptsize
\centering
\begin{tabular}{>{\centering}p{3mm}>{\centering}p{3mm}>{\centering}p{1mm}>{\centering}p{3mm}>{\centering}p{1mm}>{\centering}p{3mm}>{\centering}p{5mm}>{\centering}p{3mm}}

\cline{4-4}

{} &
{} &
\multicolumn{1}{c}{} & 
\multicolumn{1}{|c|}{\multirow{2}{*}{\large{$ip\textsubscript{0}$}}} &
\multicolumn{1}{c}{} &
\multicolumn{1}{c}{} &
\multicolumn{1}{c}{} & 
\multicolumn{1}{c}{}\cr

\hhline{|-|-|~|~|~|-|~|~|}

\multicolumn{1}{|c|}{\cellcolor{dgray}{\large{$wt\textsubscript{0}$}}} &
\multicolumn{1}{|c|}{\cellcolor{dgray}{\large{$wt\textsubscript{1}$}}} &
\multicolumn{1}{c}{\multirow{2}{*}{\large{$\times$}}} & 
\multicolumn{1}{|c|}{} & 
\multicolumn{1}{c}{\multirow{2}{*}{\large{$+$}}} & 
\multicolumn{1}{|c|}{\cellcolor{dgray}{\large{$b$}}} &
\multicolumn{1}{c}{\normalsize{\text{ReLU}}} & 
\multicolumn{1}{c}{\multirow{2}{*}{\large{$out$}}} \cr

\cline{4-4}

\multicolumn{1}{|c|}{\normalsize{$1$}} &
\multicolumn{1}{c|}{\normalsize{$-1$}} & 
\multicolumn{1}{c}{} & 
\multicolumn{1}{|c|}{\multirow{2}{*}{\large{$ip\textsubscript{1}$}}} &
\multicolumn{1}{c}{} &
\multicolumn{1}{|c|}{\normalsize{$-33$}} &
\multicolumn{1}{c}{\large{$\longrightarrow$}} & 
\multicolumn{1}{c}{} \cr

\cline{1-2}\cline{6-6}

{} &
{} & 
\multicolumn{1}{c}{} & 
\multicolumn{1}{|c|}{} & 
\multicolumn{1}{c}{} &
\multicolumn{1}{c}{} &
\multicolumn{1}{c}{} & 
\multicolumn{1}{c}{}\cr

\cline{4-4}

\end{tabular}
\captionof{figure}{Example BNN-based neuron for demonstrating the model recovery attacks.} 
\label{fig:bnn_example}
\end{table}
We will now demonstrate our methodology of the model recovery using the example neuron of Fig. \ref{fig:bnn_example}:

\textbf{Step I. Extraction of weights ($wt$):} The multiplication operation comprises of either directly passing $ip$ or conditional negating it (depending on whether $wt$ is $+ / -$ 1 respectively) and then adding it to $pa$. This conditional negation of $ip$ (for $wt=-1$) leads to more number of cycles and thus, produces a timing side-channel. A major advantage is that all the weights $wt$'s can be extracted in parallel using this information.

\textbf{Step II. Extraction of bias ($b$):} After extracting, the weights $wt$'s, we initialize the inputs such that $pa$ is minimized. For our example neuron of Fig. \ref{fig:bnn_example},  $ip\textsubscript{0}$ is set to $0$ and $ip\textsubscript{1}$ is set to $255$. We then increment $pa$ by the smallest quanta by appropriately setting the inputs, till $pa$ changes its sign. This change in sign is observed by change in the timing of the ReLU operation (same as that of the fixed point ReLU operation in Fig. \ref{fig:fxp_relu}). For our example neuron, we perform this operation by decreasing $ip\textsubscript{1}$ to $0$ while keeping $ip\textsubscript{0}=0$ and then increasing $ip\textsubscript{0}$ gradually.  For our example neuron, we find that $pa$ changes its sign when $ip\textsubscript{0}$ is $33$ and $ip\textsubscript{1}$ is $0$. Thus $b = -\left( wt\textsubscript{0} \times 33 + wt\textsubscript{1} \times 0\right)  = -33$. 

\textbf{Step III. Extension to successive layers:} We proceed successively layer-wise to recover the weights ($wt$'s) and biases ($b$'s) using Steps I and II respectively.

We validated our attack methodology on a real 2-layer perceptron neural network for MNIST digit recognition, with different bit precisions adapted for floating point, fixed point and binary networks. For floating point, all the weights and bias were recovered with $< 1\%$ error. Exact model recovery was achieved for both fixed point and binary networks.

\section{Input Recovery}\label{sec:input_recovery}
We  will now  discuss  the recovery of  inputs  for floating point (Section \ref{sec:input_flp}) and normalization-based (Section \ref{sec:input_nrm}) NNs. Extraction of sparse inputs (e.g. MNIST \cite{mnist_dataset}) for zero-skipping-based NNs is also discussed in Section \ref{sec:input_pnn}. We will demonstrate our methodology for common image recognition applications which use 8-bit unsigned integers. It is important to reiterate that for input recovery, the attacker is assumed to have pre-characterized the hardware platform.


\subsection{Input recovery for floating point neural networks} \label{sec:input_flp} 
Any 8-bit unsigned integer input $ip$ can be represented as an equivalent floating point number with its mantissa and exponent as shown below in eq. \ref{eqn_int2float}. 
\begin{equation} \label{eqn_int2float}
    ip=1.m\textsubscript{ip}\times2\textsuperscript{e\textsubscript{ip}},1.m\textsubscript{ip}\in\normalsize{\left\{1,...,1\frac{127}{128}\right\},e\textsubscript{ip}\in\{0,...,7\}}
\end{equation}

\textbf{Step I. Extraction of mantissa ($1.m\textsubscript{ip}$):} The timing side-channel of floating point multiplication was used to recover the mantissa of the weights in Section \ref{sec:model_flp}.  We adopt the same approach for extracting the input mantissa by constructing a LUT of $1.m\textsubscript{wt} \xrightarrow {1.m\textsubscript{\tiny{ip}}} T(1.m\textsubscript{ip}\times1.m\textsubscript{wt})$ mapping for all the possible input activation mantissas $1.m\textsubscript{ip}$ and all unique $1.m\textsubscript{wt}$'s in the 1$\textsuperscript{st}$ layer (that directly accepts the inputs). For an unknown input $ip\textsubscript{0}$, we obtain its corresponding $1.m\textsubscript{wt} \xrightarrow{1.m\textsubscript{ip\textsubscript{0}}}T(1.m\textsubscript{ip\textsubscript{0}}\times1.m\textsubscript{wt})$ mapping and then match it with the pre-characterized LUT to determine $1.m\textsubscript{ip\textsubscript{0}}$.

The requirement of no prior knowledge about $wt$ / $1.m\textsubscript{wt}$ is one of the major advantage of this method. The weights can be known or can be extracted using the method discussed in Section \ref{sec:model_flp}. The accuracy of mantissa extraction depends on the number of unique weight mantissas involved. All current NNs use more than sufficient number of unique mantissas in the first layer to recover all $1.m\textsubscript{ip}$ values exactly.

\textbf{Step II. Extraction of exponent ($e\textsubscript{ip}$):} For floating point multiplication operation, where one operand $ip$ is an integer (e.g., obtained from sensor), $ip$ is first converted to  floating point number and then multiplied using the conventional floating point multiplication. For the ATmega328P platform, the timing of the integer-to-float conversion can be used to recover the exponent of the input $ip$, as shown in Fig. \ref{fig:int2float_time}. Please refer to the detailed analysis of the integer-to-float conversion process in Appendix \ref{app:int2float}.



\begin{figure}[!ht]
        \centering
        {\includegraphics[width=\linewidth]{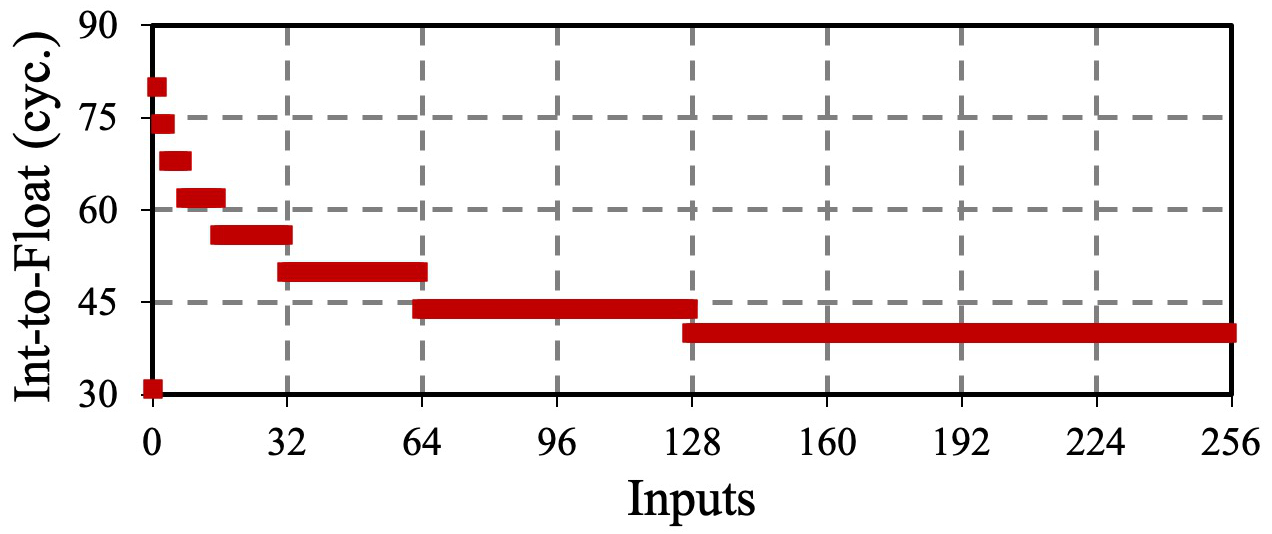}}
        \caption{Timing side-channel of the integer-to-float conversion is used to recover the equivalent exponent $e\textsubscript{ip}$.}
        \label{fig:int2float_time}
\end{figure}


\subsection{Input recovery using normalization operation}\label{sec:input_nrm}

\begin{figure}[!ht]
        \centering
        {\includegraphics[width=\linewidth]{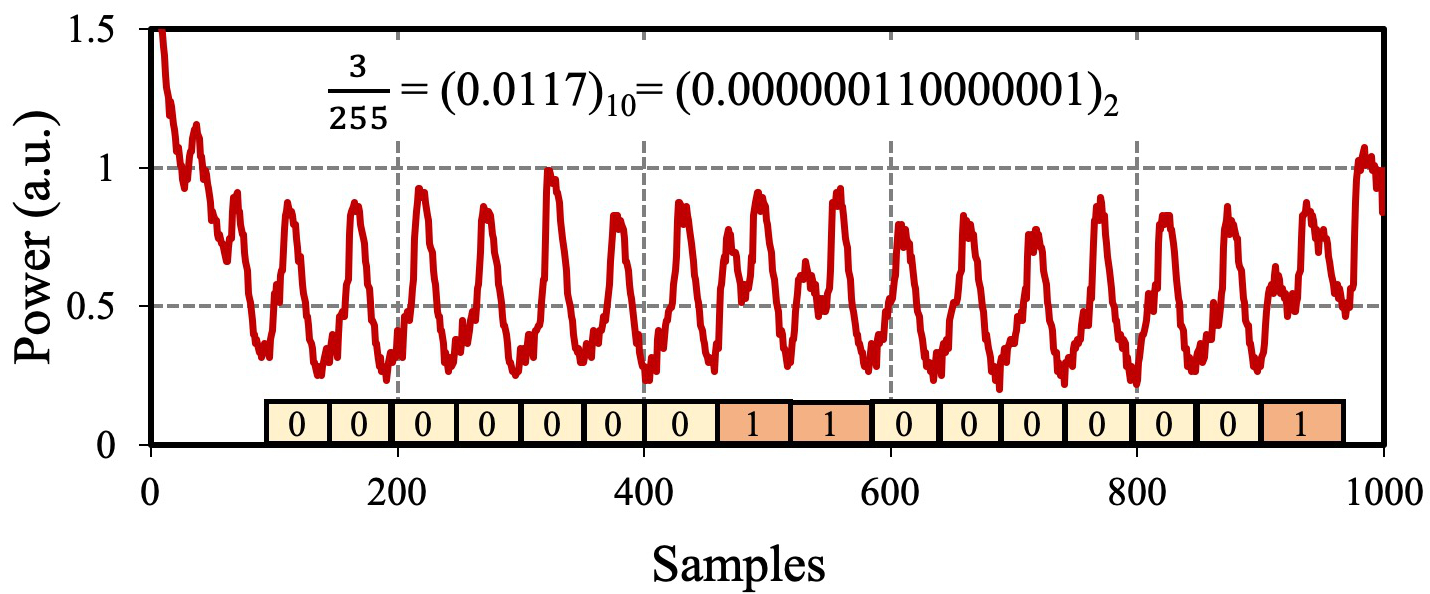}}
        \caption{Input recovery for fixed point normalization operation using SPA attack over the power waveform.}
        \label{fig:fxp_input_normalization}
\end{figure}

We propose a SPA-based input recovery method that is \textit{strictly applicable} when the inputs are normalized using division operation (sometimes used during pre-processing of data).
In our demonstration, the input $ip$ is normalized to [0,1] using $\frac{ip}{255}$ and the output $op$ is stored as a 16-bit fixed point representation $(op\textsubscript{0}\cdot op\textsubscript{-1}...op\textsubscript{-15})$. This division operation involves a series of conditional subtractions depending on the exact bit sequence of the output. Only when the quotient bit is 1, the divisor is subtracted from the dividend (as commonly done in conventional division operation). As shown in Fig. \ref{fig:fxp_input_normalization}, bits corresponding to $1$ in $op$ give rise to a double peak in the power trace and consume more timing, which can be used to recover $ip$. 

Fig. \ref{fig:image_recovery} shows the inputs recovered using the proposed techniques for ImageNet (``red car'' and ``dog''), CIFAR10 (``horse'' and ``ship'') and MNIST (``digit 0'') dataset. Exact input recovery was achieved in all cases for both (b) and (c).

\begin{table}[!htp]
\scriptsize
\centering
\begin{tabular}{>{\centering}p{0.30\linewidth}>{\centering}p{0.30\linewidth}>{\centering}p{0.30\linewidth}}
{\multirow{8}{*}{\includegraphics[width=0.95\linewidth]{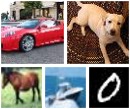}}} & {\multirow{8}{*}{\includegraphics[width=0.95\linewidth]{Figures/image_recovery.jpg}}} &
{\multirow{8}{*}{\includegraphics[width=0.95\linewidth]{Figures/image_recovery.jpg}}} \cr
{} & {} & {} \cr
{} & {} & {} \cr
{} & {} & {} \cr
{} & {} & {} \cr
{} & {} & {} \cr
{} & {} & {} \cr
{} & {} & {} \cr
{\fontsize{8}{12}\selectfont $\text{(a)}$} & {\fontsize{8}{12}\selectfont $\text{(b)}$} & {\fontsize{8}{12}\selectfont $\text{(c)}$} \cr
\end{tabular}
\captionof{figure}{(a) Actual inputs and recovered inputs using (b) floating point NN-based input recovery (Section \ref{sec:input_flp}) (c) normalization-based input recovery (Section \ref{sec:input_nrm}).}
\label{fig:image_recovery}
\end{table}

\subsection{Sparse input recovery for zero-skipping neural networks}\label{sec:input_pnn}
For high energy-efficiency, zero-skipping-based neural networks do not execute any multiplication when either of the operands is 0 \cite{2016_han}. For $ip= 0$, the corresponding weights are not fetched from the memory for multiplication. This timing helps to recover specialized sparse inputs like MNIST reasonably well by separating zero v/s non-zero inputs (Fig. \ref{fig:pruned_image_recovery}). However, this method cannot be applied to general images.

\begin{table}[!ht]
\scriptsize
\centering
\begin{tabular}{c}
{\multirow{2}{*}{\includegraphics[width=0.9\linewidth]{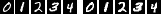}}} \cr
{} \cr
{} \cr
{\fontsize{9}{12}\selectfont $\text{\hspace{0.1cm} (a) Original MNIST inputs \hspace{0.1cm} (b) Recovered MNIST inputs}$} \cr
\end{tabular}
\captionof{figure}{MNIST input recovery for zero-skipping-based NN.}
\label{fig:pruned_image_recovery}
\end{table}
\begin{figure*}[!htp]
\centering
\includegraphics[width=\linewidth]{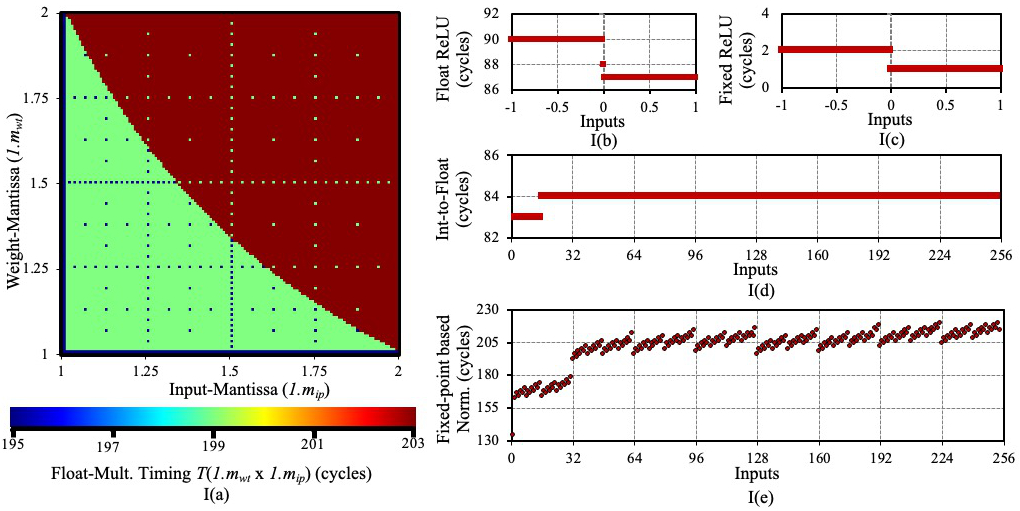}
\includegraphics[width=\linewidth]{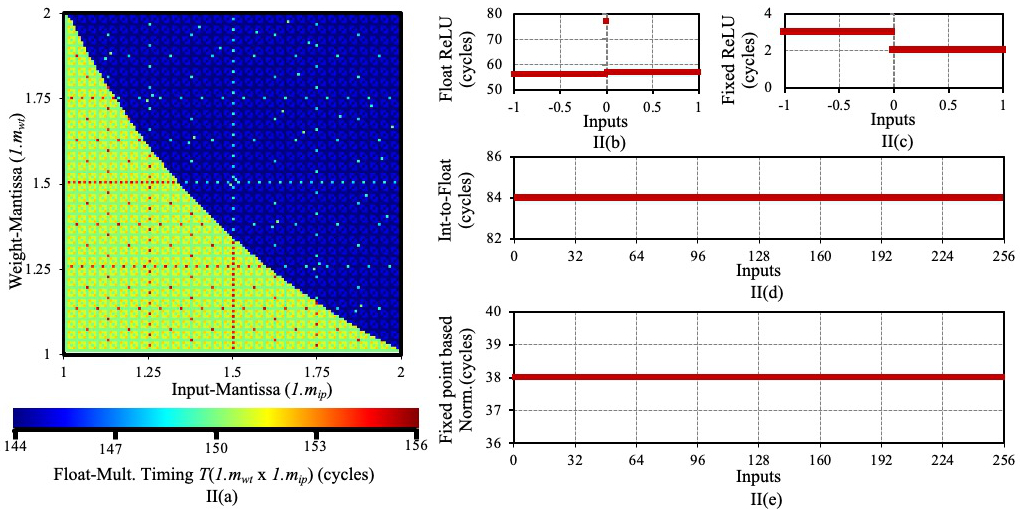}
\caption{Side-channel leakages from (I) ARM Cortex-M0+ and (II) RISC-V RV32IM micro-processors: (a) floating point multiplication timing (input and weight parameters same as Fig. \ref{fig:flp_mantissa_lut}); Timing side-channel of (b) floating point ReLU, (c) fixed point ReLU, (d) integer-to-float conversion and (e) fixed point normalization operation.}
\label{fig:other_platform}.
\end{figure*}

\begin{table*}[!htp]
\renewcommand{\arraystretch}{1.1}
\scriptsize
\centering
\captionof{table}{\footnotesize \MakeTextUppercase{Specifications of Embedded Micro-Processors used for our Experimental Demonstration}}
\begin{tabular}{|>{\centering}p{30mm}|>{\centering}p{14mm}|>{\centering}p{20mm}|>{\centering}p{20mm}|>{\centering}p{22mm}|>{\centering}p{20mm}|} \hline

\rowcolor{lgray}
\textbf{Processor} & \textbf{Architecture} & \textbf{Addition} & \textbf{Multiplication} & \textbf{Division} & \textbf{Floating point} \cr
\hline
\multirow{2}{*}{Atmel ATmega328P \cite{microchip_atmega, arduino_uno}} & \multirow{2}{*}{8-bit} & {Hardware} & {Hardware } & \multirow{2}{*}{Software} & \multirow{2}{*}{Software}  \cr 
& & (8-bit + 8-bit) & (8-bit $\times$ 8-bit) & & \cr \hline
\multirow{2}{*}{ARM Cortex-M0+ \cite{microchip_arm_cm0plus, adafruit_metro_m0_express}} & \multirow{2}{*}{32-bit} & {Hardware} & {Hardware } & \multirow{2}{*}{Software} & \multirow{2}{*}{Software}  \cr 
& & (32-bit + 32-bit) & (32-bit $\times$ 32-bit) & & \cr \hline

\multirow{2}{*}{RISC-V RV32IM \cite{banerjee_jssc_2019,banerjee_tches_2019}} & \multirow{2}{*}{32-bit} & {Hardware} & {Hardware } & {Hardware} & \multirow{2}{*}{Software} \cr 
& & (32-bit + 32-bit) & (32-bit $\times$ 32-bit) & {(32-bit / 32-bit}) & \cr \hline

\end{tabular}
\label{tab:specifications}
\end{table*}

\section{Extension to Other Platforms}\label{sec:other_platform}
Next, we discuss the extensions of our previously discussed attack techniques to the following embedded platforms:

\textbf{ARM Cortex-M0+:} We demonstrate the attacks on the ATSAMD21G18 ARM Cortex-M0+ micro-controller \cite{microchip_arm_cm0plus} available in the Adafruit Metro M0 Express board \cite{adafruit_metro_m0_express}. It has a 32-bit architecture with a single-cycle 32-bit $\times$ 32-bit hardware multiplier. It does not have a hardware floating point unit (FPU) and supports only software implementations of floating point arithmetic.

\textbf{RISC-V RV32IM:} We also demonstrate our attacks on a custom RISC-V micro-processor chip supporting the RV32IM instruction set \cite{banerjee_jssc_2019, banerjee_tches_2019}. This processor has no hardware FPU and performs floating point operations in software. However, apart from the 1-cycle hardware multiplier, it also has a 32-cycle constant-time hardware divider, as the RV32IM instruction set includes both multiplication and division instructions.

Figs. \ref{fig:other_platform} (I) and (II) display the relevant timing side-channel leakages from ARM Cortex-M0+ and RISC-V processors respectively. Table \ref{tab:specifications} also presents the specifications of the three micro-processors used for our experimental evaluations. The major observations are summarized below:

\begin{itemize}
    \item The floating point multiplication operation timings for both Cortex-M0+ and RISC-V (Fig. \ref{fig:other_platform}:I-II(a)) show first-order trends and hence, allows an attacker to retrieve the mantissa information of the weights, as was analyzed in Section \ref{sec:model_flp}.
    \item The floating point ReLU (Fig. \ref{fig:other_platform}:I-II(b)) and fixed point ReLU (Fig. \ref{fig:other_platform}:I-II(c)) operations have timing side-channels differentiating positive and negative inputs. This can be used to identify zero-crossing points and hence, are in agreement with the previously described attacks (Section \ref{sec:model_flp} and \ref{sec:model_fxp}).
    \item The integer-to-float conversion process of Cortex-M0+ (Fig. \ref{fig:other_platform}:I(d)) has negligible information leakage. Thus, the values of the exponent cannot be determined from this timing information. Also, the floating point multiplication timing (Fig. \ref{fig:other_platform}:I(a)) has fewer variations and hence, a lot of weights are required to identify the input mantissas precisely. Thus, the input recovery for Cortex-M0+ is difficult to perform. The integer-to-float conversion process for RISC-V (Fig. \ref{fig:other_platform}:II(d)) takes constant time (irrespective of the input), hence the exponent information cannot be recovered. Hence, the extraction of inputs for floating point-based NNs is also difficult to perform on the RISC-V platform  
    \item In case of Cortex-M0+, the fixed point division operation, in software, has timing proportional to hamming weight of the output (Fig.\ref{fig:other_platform}:I(e)), and hence can be used to recover the inputs. The custom RISC-V chip has a 32-cycle hardware divider, because of which the normalization operation takes constant time (Fig. \ref{fig:other_platform}:II(e)). Thus, normalization-based input recovery attacks (Section \ref{sec:input_nrm}) can be prevented by having a constant-time dedicated hardware divider, albeit at the cost of increased logic area in the chip.
\end{itemize}

\section{Proposed Countermeasures}\label{sec:defense}
Finally, we propose countermeasures against the previously discussed attacks and analyze their implementation overheads.

\textbf{Floating point MAC operation:} The primary reason for the timing side-channel leakage of floating point-based multiplication operation is that the output of every stage is represented according to the IEEE-754 representation \cite{2019_ieee754}. So, instead of representing the numbers according to the conventional IEEE-754 representation, we alternatively represent them by equalizing the exponents with respect to the maximum exponent $e\textsubscript{max}$ in the layer and storing only the modified mantissa information along with the signs. Thus, any generic floating point weight $wt$ is modified to be represented in fixed point as: $wt = (wt\textsubscript{23},wt\textsubscript{22},...,wt\textsubscript{0})\textsubscript{2} = (-1)\textsuperscript{s\textsubscript{wt}}\times1.m\textsubscript{wt}\times 2\textsuperscript{e\textsubscript{wt}-e\textsubscript{max}}$, which requires storage of only 3 bytes (instead of 4 bytes). For example, if the $wt$'s of a layer are $1.75\times2\textsuperscript{0}$, $-1.32\times2\textsuperscript{-1}$ and $1\times2\textsuperscript{-2}$, we normalize them by 2$\textsuperscript{0}$ and store the weights as $1.75, -0.66$ and $0.25$ respectively. The input activations ($ip$), similar to weights ($wt$), are also normalized. 



TABLE \ref{tab:float_defense} shows the overheads for our proposed defense for floating point operations on different platforms. The timing of the multiplication operation increased by $\sim2\times$ for ATmega328P. However, the timing of an ensemble of 25 MAC operations didn't show a proportional increase because the  addition operations are now simplified to traditional fixed point signed additions. 

\textbf{ReLU operation:} We propose the following constant-time implementation of the ReLU operation for 8-bit input $pa$:
\begin{equation}
    {\begin{myfont} \normalsize{\text{out} = (\sim(\text{pa\hspace{2mm}>>\hspace{2mm}7}))\text{\hspace{2mm}\&\hspace{2mm}pa}}\end{myfont}}\notag
\end{equation}
Here, the sign bit (most significant bit) of $pa$ is extracted, right-shifted and inverted to create a mask. The mask consists of all 0's for negative inputs and all 1's for positive inputs. This mask is AND-ed with original $pa$ to compute the final output $out$. Our proposed ReLU operation for 8-bit integer operands is shown in TABLE \ref{tab:relu_defense} along with the intermediate values of computation. None of the intermediate steps of our proposed method is data-dependent, and hence not susceptible to timing attacks. Our proposed method is applicable to both fixed point and floating point operands (using our previously proposed representation). TABLE \ref{tab:float_defense} show the timings of our proposed ReLU implementation for 16-bit fixed point and 32-bit floating point inputs respectively. The additional steps lead to increase in cycle count for our proposed constant-time fixed point ReLU operation. However, the performance improves for floating point ReLU because the expensive floating point-based comparison operation is eliminated. 

\begin{table}[!hb] 
\centering
\captionof{table}{\footnotesize \MakeTextUppercase{Proposed constant-time ReLU for 8-bit integer operands (exemplified by +ve and -ve inputs)}}
\begin{tabular}{|>{\centering}p{18mm}|>{\centering}p{24mm}|>{\centering}p{24mm}|} 
\hline
\rowcolor{lgray}\multicolumn{3}{|c|}{} \cr
\rowcolor{lgray}
\multicolumn{3}{|c|}{\multirow{-2}{*}{\normalsize\textbf{{\begin{myfont}out =  ($\sim$(pa >> 7))\hspace{2mm}\&\hspace{2mm}pa \end{myfont}}}}}\cr
\hline

\multirow{3}{*}{\normalsize{\begin{myfont}pa\end{myfont}}} &
\multirow{2}{*}{\normalsize{\begin{myfont}-123$\textsubscript{10}$\end{myfont}}} &
\multirow{2}{*}{\normalsize{\begin{myfont}123$\textsubscript{10}$\end{myfont}}} \cr
&\multirow{2}{*}{\normalsize{\begin{myfont}10000101$\textsubscript{2}$\end{myfont}}} 
&\multirow{2}{*}{\normalsize{\begin{myfont}01111011$\textsubscript{2}$\end{myfont}}} \cr
& & \cr
\hline
\multirow{2}{*}{\normalsize{\begin{myfont}$\sim$(pa>>7)\end{myfont}}} & \multirow{2}{*}{\normalsize{\begin{myfont}00000000$\textsubscript{2}$\end{myfont}}} &
\multirow{2}{*}{\normalsize{\begin{myfont}11111111$\textsubscript{2}$\end{myfont}}} \cr
& & \cr
\hline
\multirow{2}{*}{\normalsize{\begin{myfont}out\end{myfont}}} & 
\multirow{2}{*}{\normalsize{\begin{myfont}00000000$\textsubscript{2}$\end{myfont}}} &
\multirow{2}{*}{\normalsize{\begin{myfont}01111011$\textsubscript{2}$\end{myfont}}} \cr
& & \cr
\hline
\end{tabular}
\label{tab:relu_defense}
\end{table}

\begin{table*}[!ht]
\renewcommand{\arraystretch}{1.2}
\scriptsize
\centering
\captionof{table}{\footnotesize \MakeTextUppercase{Performance Analysis of Side-channel countermeasures For Common NN operations}}
\begin{tabular}{|p{16mm}|p{14mm}|>{\centering}p{16mm}|>{\centering}p{16mm}|>{\centering}p{16mm}|>{\centering}p{16mm}|>{\centering}p{16mm}|>{\centering}p{16mm}|} \hline

\multicolumn{2}{|c|}{} &
\multicolumn{2}{ c|}{\cellcolor{lgray}{\textbf{Atmel ATmega328P}}} & 
\multicolumn{2}{ c|}{\cellcolor{white}{\textbf{ARM Cortex-M0+}}} &
\multicolumn{2}{ c|}{\cellcolor{lgray}{\textbf{RISC-V RV32IM}}}  \cr
\hhline{|~|~|-|-|-|-|-|-|}

\multicolumn{2}{|c|}{\multirow{-2}{*}{\textbf{Metric}}}  & \cellcolor{lgray}{Default} & \cellcolor{lgray}{Solution} & 
\cellcolor{white}{Default} & \cellcolor{white}{Solution} &
\cellcolor{lgray}{Default} & \cellcolor{lgray}{Solution} \cr \hline

\multicolumn{2}{|l|}{Cycles for multiplication op.} & \cellcolor{lgray}{147.6 (avg.)} & \cellcolor{lgray}{300} & \cellcolor{white}{201.3 (avg.)} & \cellcolor{white}{ 59} &
\cellcolor{lgray}{146.9 (avg.)} & \cellcolor{lgray}{ 8} 
\cr \hline

\multicolumn{2}{|l|}{Cycles for 25 MAC op.\textsuperscript{$\dagger$}} & \cellcolor{lgray}{6828} & \cellcolor{lgray}{8541} & \cellcolor{white}{8406} & \cellcolor{white}{2258} &
\cellcolor{lgray}{7530} & \cellcolor{lgray}{209} 
\cr\hline

\multicolumn{2}{|l|}{Storage space for weights} & \cellcolor{lgray}{1$\times$}& \cellcolor{lgray}{0.75$\times$}& \cellcolor{white}{1$\times$}& \cellcolor{white}{0.75$\times$}&
\cellcolor{lgray}{1$\times$}& \cellcolor{lgray}{0.75$\times$}
\cr \hline

\multicolumn{1}{|l|}{Cycles of ReLU} & {+ve input} & \cellcolor{lgray}{68} & \cellcolor{lgray}{36} & \cellcolor{white}{90} & \cellcolor{white}{10} &
\cellcolor{lgray}{56} & \cellcolor{lgray}{6} \cr

\multicolumn{1}{|l|}{(floating point)} & {{0 } { input}} & 
\cellcolor{lgray}{56} & \cellcolor{lgray}{(constant time} & \cellcolor{white}{88} & \cellcolor{white}{(constant time} &
\cellcolor{lgray}{77} & \cellcolor{lgray}{(constant time} \cr

{} & {-ve input} & 
\cellcolor{lgray}{61} & \cellcolor{lgray}{for any input)} & \cellcolor{white}{87} & \cellcolor{white}{for any input)} &
\cellcolor{lgray}{57} & \cellcolor{lgray}{for any input)}
\cr \hline

\multicolumn{1}{|l|}{Cycles of ReLU} & {+ve input} 
& \cellcolor{lgray}{10} & \cellcolor{lgray}{19} 
& \cellcolor{white}{2}  & \cellcolor{white}{9}   
& \cellcolor{lgray}{3}  & \cellcolor{lgray}{6}   \cr 

\multicolumn{1}{|l|}{(fixed point)} & {{0 } { input}} & 
\cellcolor{lgray}{7} & \cellcolor{lgray}{(constant time} & \cellcolor{white}{1} & \cellcolor{white}{(constant time} &
\cellcolor{lgray}{2} & \cellcolor{lgray}{(constant time} \cr 

& {-ve input} & 
\cellcolor{lgray}{7} & \cellcolor{lgray}{for any input)} & 
\cellcolor{white}{1} & \cellcolor{white}{for any input)} & 
\cellcolor{lgray}{2} & \cellcolor{lgray}{for any input)} \cr \hline

\multicolumn{8}{l}{\textsuperscript{$\dagger$} The weights are uniformly chosen from (-1,1) and the inputs are chosen from (0,1). }           \cr

\multicolumn{8}{l}{\textbf{\textit{Note:}} All the operation-specific cycles measurements reported in this paper may incur few extra cycles because of reading/writing of data.} \cr

\end{tabular}
\label{tab:float_defense}
\end{table*}

\begin{figure}[!ht]
\centering
\includegraphics[width=\linewidth]{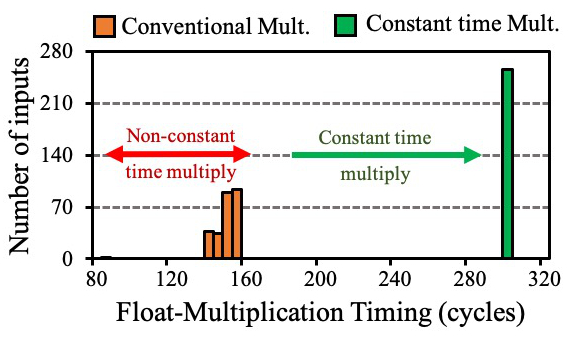}
\caption{Histogram for the cycle counts of conventional and proposed floating point-based multiplication operation of $ip\in\{0,...255\}$ with $wt$=2.34 for ATmega328P.}
\label{fig:input_leakage}.
\end{figure}

\textbf{Input recovery:} Our proposed normalized representation for floating point operands removes any timing side-channel during MAC operations. Therefore, the previously described attacks (Section \ref{sec:input_flp}) for extracting the equivalent mantissa information is no longer applicable. This is better elucidated in Fig. \ref{fig:input_leakage}, which plots the histogram of the inputs with respect to the cycle counts for conventional and proposed multiplication operation ($ip\in\{0,...255\}$ and $wt=2.34$) for ATmega328P. The variation in the timing of the conventional multiplication operation for constant $wt$ leaks some information about $ip$. However, our proposed constant-time multiplication removes any timing-based leakage of inputs. Similarly, for NN with division-based input normalization, the division operation can be completely eliminated by appropriately scaling the relevant parameters of the first layer of the NN. This allows the NN to directly take in the raw inputs without the need for normalization.


\section{Future Work}\label{sec:future_work}

We now discuss some future extensions of this work:

\textbf{I. Extension to statistical attacks:} In this paper, our primary focus was to recover the model parameters of NN using only timing side-channels and SPA. While the proposed countermeasures provide constant-time operation, they may not necessarily defend against other attacks like differential power analysis (DPA) and correlation power analysis (CPA).

\begin{figure}[!ht]
    \centering
    {\includegraphics[width=0.95\linewidth]{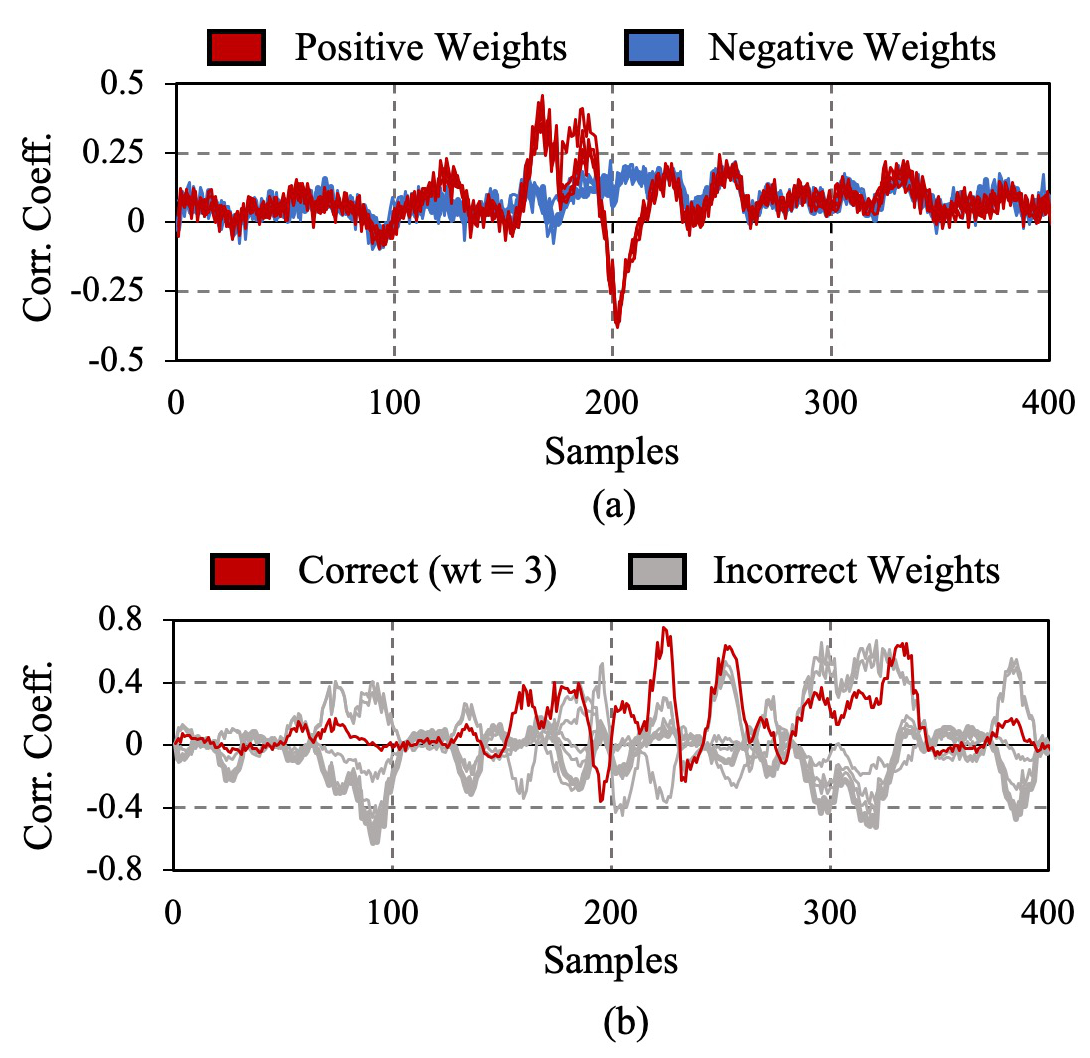}}
    \caption{CPA of constant-time fixed point multiplication based on: (a) sign and (b) magnitude of the weight.}
    \label{fig:cpa_fixed_mul}
\end{figure}

$(i)$ We performed CPA over the proposed constant-time fixed point multiplication operation on the Arduino ATmega328P platform. Unsigned integer inputs $ip\in\{0,1,...,255\}$ were provided and the byte-wise hamming weight of $ip \times wt$ model was used for the CPA. Fig. \ref{fig:cpa_fixed_mul}(a) demonstrates that CPA over the most significant byte of the output $ip\times wt$ was successful in identifying the sign of $wt$. Similarly, Fig. \ref{fig:cpa_fixed_mul}(b) shows that CPA over the least significant byte of the output $ip\times wt$ was successful in revealing $|wt|$. \cite{batina_model_2019,batina_input_2019} describes the use of CPA and statistical techniques to recover complete neural networks.

\begin{figure}[!ht]
    \centering
    {\includegraphics[width=0.95\linewidth]{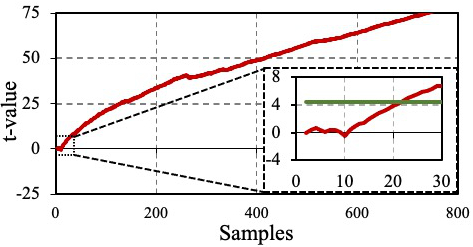}}
    \caption{Leakage assessment $t$-test for constant-time ReLU. The crossing of the t-value threshold is shown in the inset.}
    \label{fig:relu_ttest}
\end{figure}

$(ii)$ Even though the ReLU operation has been made constant time (TABLE \ref{tab:relu_defense}), the mask $\begin{myfont}\sim(pa>>7)\end{myfont}$ consists of all $0/1$'s for $-/+$ve inputs, thus leaking some information using hamming weight (and possibly power). The t-test result over the proposed ReLU operation is shown in Fig. \ref{fig:relu_ttest}. The two groups A and B contain power waveforms corresponding to randomly selected +ve and -ve inputs respectively. We get $|t| > 4.5$ for more than 25 measurements, which proves that these two groups have distinct power signatures.
 
Another viable direction to explore is the use of algebraic side-channel analytic techniques \cite{mangard_spa,schramm_sca_collision,bogdanov_algrebraic_sidechannel}. In contrast to the conventional divide-and-conquer / statistical techniques (e.g., CPA, DPA), the algebraic techniques require minimum data complexity at the cost of more complex and sensitive computational steps. Thus, it seems that algebraic techniques might be a better choice for recovering the model parameters and statistical techniques are more suited for the input recovery attacks (real-time operation with minimal complexity at the edge nodes). The use of hybrid techniques like soft analytical side-channel attacks (SASCA) \cite{sasca_xavier} also seems to be promising as it combines the low time-memory complexity and noise tolerance of standard DPA with the optimal data complexity of algebraic side-channel attacks. All the common operations execute in multiple cycles and thus, involve many intermediate steps. The dependencies between these intermediate variables make them suitable for algebraic attacks.


\textbf{II. Extension to hardware platforms:}  The attacks which were described in this paper assume that the timings of individual NN operations can be extracted. While this condition is relatively easy to meet on embedded micro-controllers, it is very difficult to achieve in the case of custom-designed hardware units such as FPGA or ASIC platforms. For improved throughput and energy-efficiency in such hardware accelerators, the common NN operations (e.g., MAC, ReLU) are executed in parallel, using an array of processing elements (PEs). In this situation, it will be difficult to obtain the timing and power corresponding to individual operations. So, the methods described in this paper may not be directly applicable to FPGA and ASIC implementations. However, following a similar methodology with larger lookup tables and crafted inputs, it may still be possible to attack such custom hardware accelerators, and this will be explored in the future.


\section{Conclusion}
In this work, we have demonstrated the recovery of model parameters and inputs for common neural network (NN) implementations with different precisions, such as floating point, fixed point and binary, on three different embedded micro-controller platforms (Atmel ATmega328P, ARM Cortex-M0+ and RISC-V RV32IM) using only timing and simple power analysis side-channel attacks.
Timing side-channel leakage from the multiplication and non-linear activation function computations was utilized to recover model parameters of floating point NNs.
For fixed point and binary NNs, zero-crossover input information obtained from the timing side-channel of the non-linear activation function was used with strategically crafted inputs for model recovery.
The inputs of floating point NNs were recovered using timing side-channel information from integer-to-float conversion and multiplication operations.
Input recovery for fixed point NNs with input normalization was performed using simple power analysis (SPA) attack on the division operation.
We have also proposed software countermeasures against these side-channel attacks and analyzed their implementation overheads. 

The feasibility and simplicity of our attacks, with minimal storage and computation requirements, emphasize the side-channel security concerns of embedded neural network implementations and the need for defending against them. The performance overheads of our proposed software-based countermeasures also motivate the design of custom side-channel-resistant hardware for embedded neural network accelerators.

\section*{Acknowledgment}
The authors acknowledge the funding support from Analog Devices and Texas Instruments. The authors sincerely thank Dr. Samuel H. Fuller, Prof. Vivienne Sze, Maitreyi Ashok and Kyungmi Lee for their valuable suggestions. The authors are also thankful to the editors and the anonymous reviewers for their insightful comments which helped improve the quality of the paper. 

\normalsize
\appendix
\section{Appendix}
\subsection{Extended analysis of floating point model extraction} \label{app:flp_wt}
(1) Derivation for 1$^{st}$ round of weight extraction:\\
Rewriting eq. \ref{eq_flp_crossover1} for $wt\textsubscript{ref}$, we obtain eq. \ref{eq_flp_crossover1_ref}.
\begin{equation} \label{eq_flp_crossover1_ref} 
    \left(ip\stackanchor{\scriptsize{(1)}}{\scriptsize{ref}} \times wt\textsubscript{ref} +b\right)\approx0 \Rightarrow ip\stackanchor{\scriptsize{(1)}}{\scriptsize{ref}}\approx
    \frac{-b}{wt\textsubscript{ref}} 
\end{equation}

Dividing eq. \ref{eq_flp_crossover1} by eq. \ref{eq_flp_crossover1_ref}, expressing $wt$ in terms of mantissa, sign and exponent and using the fact that sgn($wt\textsubscript{k}$)$ = $sgn($wt\textsubscript{ref}$) we obtain eq. \ref{eq_flp1}-\ref{eq_flp3}.

\begin{equation} \label{eq_flp1}
\frac{ip\stackanchor{\scriptsize{(1)}}{\scriptsize{k}}}{ip\stackanchor{\scriptsize{(1})}{\scriptsize{ref}}}\approx \left( \frac{wt\textsubscript{ref}}{wt\textsubscript{k}}=\frac{1.m\textsubscript{ref}\times2\textsuperscript{e\textsubscript{ref}}}{1.m\textsubscript{k}\times2\textsuperscript{e\textsubscript{k}}} \right)
\end{equation}
\begin{equation} \label{eq_flp2}
\Rightarrow e\textsubscript{k}-e\textsubscript{ref}\approx \text{log}\textsubscript{2}\left(\frac{1.m\textsubscript{ref}}{1.m\textsubscript{k}} \times \frac{ip\stackanchor{\scriptsize{(1)}}{\scriptsize{ref}}}{ip\stackanchor{\scriptsize{(1)}}{\scriptsize{k}}}\right)
\end{equation}
\begin{equation} \label{eq_flp3}
\Rightarrow e\textsubscript{k}-e\textsubscript{ref}=\left\lceil \text{log}\textsubscript{2}\left(\frac{1.m\textsubscript{ref}}{1.m\textsubscript{k}} \times \frac{ip\stackanchor{\scriptsize{(1)}}{\scriptsize{ref}}}{ip\stackanchor{\scriptsize{(1)}}{\scriptsize{k}}}\right)\right\rfloor
\end{equation}
Here $\lceil{x}\rfloor$ round of $x$ to the nearest integer.

(2) Derivation for 2$^{nd}$ round of weight extraction:\\
For the definition of $ip\stackanchor{\scriptsize{(2)}}{\scriptsize{k}}$ we have:
\begin{equation} \label{eq_flp4} 
    \left(ip\stackanchor{\scriptsize{(2)}}{\scriptsize{k}} \times wt\textsubscript{k} + 255 \times wt\textsubscript{ref} + b\right)\approx0 
\end{equation}
\begin{equation} \label{eq_flp5} 
\Rightarrow ip\stackanchor{\scriptsize{(2)}}{\scriptsize{k}}\approx
    \frac{-\left(b + 255\times wt\textsubscript{ref}\right)}{wt\textsubscript{k}} 
\end{equation}

Rearranging eq. \ref{eq_flp_crossover1_ref} and eq. \ref{eq_flp5} (by eliminating $b$), expressing $wt$ in terms of mantissa, sign and exponent and using the fact that sgn($wt\textsubscript{k}$)$\neq$sgn($wt\textsubscript{ref}$), we obtain eq. \ref{eq_flp6}-\ref{eq_flp7}.
\begin{equation} \label{eq_flp6}
\frac{ip\stackanchor{\scriptsize{(2)}}{\scriptsize{k}}}{\left(255-ip\stackanchor{\scriptsize{(1})}{\scriptsize{ref}}\right)}\approx \left( \frac{-wt\textsubscript{ref}}{wt\textsubscript{k}}=\frac{1.m\textsubscript{ref}\times2\textsuperscript{e\textsubscript{ref}}}{1.m\textsubscript{k}\times2\textsuperscript{e\textsubscript{k}}} \right)
\end{equation}
\begin{equation} \label{eq_flp7}
\Rightarrow e\textsubscript{k}-e\textsubscript{ref}=\left\lceil \text{log}\textsubscript{2}\left(\frac{1.m\textsubscript{ref}}{1.m\textsubscript{k}} \times \frac{\left(255-ip\stackanchor{\scriptsize{(1)}}{\scriptsize{ref}}\right)}{ip\stackanchor{\scriptsize{(2)}}{\scriptsize{k}}}\right)\right\rfloor
\end{equation}

(3) Derivation for 3$^{rd}$ round of weight extraction:\\
For the definition of $ip\stackanchor{\scriptsize{(3)}}{\scriptsize{k}}$ we have:
\begin{equation} \label{eq_flp8} 
    \left(ip\stackanchor{\scriptsize{(3)}}{\scriptsize{k}} \times wt\textsubscript{ref} + 255 \times wt\textsubscript{k} + b\right)\approx0 
\end{equation}
\begin{equation} \label{eq_flp9} 
\Rightarrow ip\stackanchor{\scriptsize{(3)}}{\scriptsize{k}}\approx
    \frac{-\left(b + 255\times wt\textsubscript{k}\right)}{wt\textsubscript{ref}} 
\end{equation}

Rearranging eq. \ref{eq_flp_crossover1_ref} and eq. \ref{eq_flp9} and expressing $wt$ in terms of mantissa, sign and exponent, we obtain eq. \ref{eq_flp10}-\ref{eq_flp12}.
\begin{equation} \label{eq_flp10}
\frac{ip\stackanchor{\scriptsize{(1})}{\scriptsize{ref}}-ip\stackanchor{\scriptsize{(3)}}{\scriptsize{k}}}{255}\approx \left( \frac{wt\textsubscript{k}}{wt\textsubscript{ref}}=\frac{(-1)\textsuperscript{sgn($wt\textsubscript{k}$)}\times1.m\textsubscript{k}\times2\textsuperscript{e\textsubscript{k}}}{(-1)\textsuperscript{sgn($wt\textsubscript{ref}$)}\times1.m\textsubscript{ref}\times2\textsuperscript{e\textsubscript{ref}}} \right)
\end{equation}
\begin{equation} \label{eq_flp11}
\Rightarrow e\textsubscript{k}-e\textsubscript{ref}=\left\lceil \text{log}\textsubscript{2}\left(\frac{1.m\textsubscript{ref}}{1.m\textsubscript{k}} \times \frac{\left\lvert ip\stackanchor{\scriptsize{(3)}}{\scriptsize{k}}-ip\stackanchor{\scriptsize{(1)}}{\scriptsize{ref}}\right\rvert}{255}\right)\right\rfloor
\end{equation}
When $\left(ip\stackanchor{\scriptsize{(1)}}{\scriptsize{ref}}-ip\stackanchor{\scriptsize{(3})}{\scriptsize{k}}\right)$ $>0$, we get sgn($wt\textsubscript{k}$)$=$sgn($wt\textsubscript{ref}$).\\
When $\left(ip\stackanchor{\scriptsize{(1)}}{\scriptsize{ref}}-ip\stackanchor{\scriptsize{(3})}{\scriptsize{k}}\right)$ $<0$, we get sgn($wt\textsubscript{k}$)$\neq$sgn($wt\textsubscript{ref}$).\\ 
Thus, using the fact sgn($wt\textsubscript{ref}$)$\neq$sgn($b$), we can write:  
\begin{equation} \label{eq_flp12}
 \text{sgn}(wt\textsubscript{k})=\text{sgn}\left(\frac{ip\stackanchor{\scriptsize{(3)}}{\scriptsize{k}}}{ip\stackanchor{\scriptsize{(1)}}{\scriptsize{ref}}}-1\right) \times \text{sgn}(b)
\end{equation}

\subsection{Analysis of integer-to-float conversion for ATmega328P}
\label{app:int2float}

\begin{algorithm}[H]
	\caption{Integer-to-float conversion} 
	\begin{algorithmic}[1]
    \Procedure{INT2FLOAT}{$ip$}\Comment{Returns $r\gets ip$ as float}
	    \State $e \gets 7$  \Comment{$e$ stores the exponent of $r$} 
	    \State $z \gets ip \gg e $ \Comment{$z$ stores the mantissa of $r$} 
	    \While {$z<1$}  \label{i2f:start_while}
	        \State $z \gets 2 \times z$
            \State $e \gets e - 1$
        \EndWhile \label{i2f:end_while}
	    \State $r \gets z\times2\textsuperscript{e}$ \Comment{$z$ stores the final exponent}
	    \State \Return{r}
    \EndProcedure  
	\end{algorithmic}
	\label{algo:i2f}
\end{algorithm}

The integer-to-float conversion process for ATmega328P is explained in the form of a pseudo-code in Algorithm \ref{algo:i2f}. Relevant comments are shown alongside using $\triangleright$ symbol.

\textit{Explanation:} For an 8-bit unsigned integer $ip$, the exponent is initialized to the maximum possible value ($7$) and the mantissa $z$ is set to $\frac{ip}{128}$. Throughout the code, it is ensured that $ip$ is equal to $z\times2^{e}$. To convert to the IEEE-754 \cite{2019_ieee754} representation, the mantissa $z$ needs to lie between $[1,2)$. Hence $z$ is left shifted ($\times$ 2) and the exponent $e$ is reduced by 1, until $z\in[1,2)$. The number of times the $\textit{while}$ loop in lines \ref{i2f:start_while}-\ref{i2f:end_while} of Algorithm \ref{algo:i2f} gets executed directly depends on the exponent, as shown in Figure \ref{fig:int2float_time}.

\balance
\bibliographystyle{IEEEtran}
\bibliography{securenn}

\end{document}